\documentclass[aps,prc,superscriptaddress,showpacs,floatfix,nofootinbib,notitlepage,twocolumn]{revtex4-1}
\usepackage{amsmath,graphicx,float,hyperref}

\def\bra{\langle}
\def\ket{\rangle}
\def\bbra{\langle\!\langle}
\def\kket{\rangle\!\rangle}
\newcommand{\trento}{T$\mathrel{\protect\raisebox{-2.1pt}{R}}$ENTo}

\begin{document}

\title{ Constraining the quadrupole deformation of atomic nuclei with relativistic nuclear collisions}

\author{Giuliano Giacalone}
\affiliation{Universit\'e Paris Saclay, CNRS, CEA, Institut de physique th\'eorique, 91191 Gif-sur-Yvette, France}
\email{giuliano.giacalone@universite-paris-saclay.fr}

\begin{abstract}
Preliminary data by the STAR collaboration at the BNL Relativistic Heavy Ion Collider shows that the elliptic flow, $v_2$, and the average transverse momentum, $\bra p_t \ket$, of final-state hadrons produced in high-multiplicity $^{238}$U+$^{238}$U collisions are negatively correlated. This observation brings experimental evidence of a significant prolate deformation, $\beta\approx 0.3$, in the colliding $^{238}$U nuclei. I show that a quantitative description of this new phenomenon can be achieved within the hydrodynamic framework of heavy-ion collisions, and that thus such kind of data in the context of high-energy nuclear experiments can help constrain the quadrupole deformation of the colliding species.
\end{abstract}

\maketitle


\section{Introduction}
\label{sec:1}

Deformation is a fundamental property of atomic nuclei, reflecting the greatly collective and correlated nature of the dynamics of nucleons within the quantum many-body system. The majority of atomic nuclei possess in fact an intrinsic deformation, most notably, an axial quadrupole, or ellipsoidal, deformation. Such deformation exists when a nucleus, described by a charge (or mass) density $\rho({\bf r})$, has a nonvanishing electric (or mass) quadrupole moment:
\begin{equation}
\label{eq:quadmoment}
    \bigl \bra Y_2^0 (\Theta, \Phi) r^2 \rho({\bf r}) \bigr \ket \neq 0 , 
\end{equation}
where  ${\bf r}=(\Theta,\Phi,r)$, $Y_2^0\propto 3\cos^2 \Theta - 1 $, and angular brackets denote an expectation value with respect to the nuclear wavefunction. The magnitude of the quadrupole deformation, i.e., the eccentricity of the nuclear body, can be roughly obtained by dividing Eq~(\ref{eq:quadmoment}) by the mean-squared nuclear radius, and is typically quantified by a dimensionless \textit{deformation parameter}, $\beta$~\cite{BM}.

In the context of \textit{low-energy} nuclear experiments, the value of $\beta$ is usually inferred from measurements of the transition probability of the electric quadrupole operator from the $0^+$ ground state to the first excited $2^+$ state~\cite{Cline:1986ik,Raman:1201zz,Shou:2014eya}. Simple formulas relating this measurable quantity to the value of $\beta$ are derived under model assumptions, and a comprehensive collection of such experimentally-inferred deformation parameters is provided in Ref.~\cite{Raman:1201zz}.  Alternatively, values of $\beta$ for essentially the whole spectrum of known nuclei have been tabulated in extensive predictions, or extrapolations of theoretical models~\cite{Hilaire:2007tmk,Moller:2015fba}. These results do not necessarily agree with the figures reported in Ref.~\cite{Raman:1201zz}.

In this paper, I argue that \textit{high-energy} nuclear experiments, or relativistic heavy-ion collisions, can help place new constraints on the values of $\beta$. In these experiments, substantial evidence of a quadrupole deformation in the colliding species has emerged only recently. At the BNL Relativistic Heavy Ion Collider (RHIC), the STAR collaboration observed unambiguous signatures of nuclear deformation by means of accurate comparisons between data in $^{197}$Au+$^{197}$Au collisions and data in $^{238}$U+$^{238}$U collisions~\cite{Adamczyk:2015obl}, the latter nuclei being much more deformed in their ground state. The same kind of observations were later made as well at the CERN Large Hadron Collider (LHC), by comparing data in collisions of spherical $^{208}$Pb nuclei with collisions of deformed $^{129}$Xe nuclei~\cite{Acharya:2018ihu,Sirunyan:2019wqp,Aad:2019xmh}.
 
 It is natural that the observables measured in heavy-ion collisions enable us to probe the geometry of the colliding nuclei.  Two nuclei colliding at relativistic energy produce a quark-gluon plasma~\cite{Shuryak:2014zxa}, the high-temperature fluidlike state of strong-interaction matter. This system is produced at rest, and is set in motion by pressure gradients which are determined by its geometry at the onset of the hydrodynamic behavior. A head-on collision between deformed nuclei can yield significant spatial asymmetry in the shape of the created medium~\cite{Kuhlman:2005ts,Filip:2009zz,Hirano:2010jg,Voloshin:2010ut,Rybczynski:2012av,Schenke:2014tga,Goldschmidt:2015kpa}, which in turn triggers an \textit{anisotropic} flow of matter towards the detectors. This anisotropy is then observed in experiments as an enhanced quadrupole component in the angular distribution of final-state hadrons, dubbed elliptic flow, $v_2$~\cite{Heinz:2013th}.

However, it was recently realized~\cite{Giacalone:2019pca} that new observables exhibiting a great sensitivity to the deformation of the colliding nuclei can be constructed by looking at the magnitude of the \textit{isotropic} flow of particles, which is given by the average momentum of hadrons, $\bra p_t \ket$, in the plane orthogonal to the collision axis. This quantity does not carry information about the spatial anisotropy of the quark-gluon plasma, but is sensitive to its size and temperature (or energy). The realization of Ref.~\cite{Giacalone:2019pca} is that systems emitting the same number of particles can be classified according to their temperature by looking at their $\bra p_t \ket$. By doing so, one can get an experimental handle on the orientation of the nuclear ellipsoids at the time of interaction. This idea is reviewed in detail in Sec.~\ref{sec:3}. The outcome is that in nearly-head-on collisions at fixed final-state multiplicity, there should exist a negative correlation between elliptic flow, $v_2$, and the average transverse momentum of hadrons, $\bra p_t \ket$, an effect which is entirely engendered by the nonzero value of $\beta$ in the colliding species. Remarkably enough, this prediction has been already investigated by the STAR collaboration at RHIC, and preliminary experimental data confirms~\cite{shengli} the existence of a negative correlation between $v_2$ and $\bra p_t \ket$ in central $^{238}$U+$^{238}$U collisions.

Here I show that this new measurement represents a crucial leap towards a quantitative phenomenology of nuclear structure at high-energy nuclear colliders. I take the idea of Ref.~\cite{Giacalone:2019pca}, and I improve its phenomenological ingredients to show that, within the hydrodynamic framework of heavy-ion collisions, it is indeed possible to achieve a quantitative description of the new phenomenon observed by the STAR collaboration. The most important feature introduced in this work is the implementation of a more realistic initial-state predictor for $\bra p_t \ket$, as I shall explain through Sec.~\ref{sec:2}. Secondly, I implement in a realistic way the statistical fluctuations that plague the determination of $\bra p_t \ket$ in a single nucleus-nucleus collision event, due to the finite hadron multiplicity, as I explain in Sec.~\ref{sec:33}. This improved setup allows me to obtain results that can be considered as quantitative predictions, as shown in Sec.~\ref{sec:4}, and that can be compared with future data. Further assumptions made in Ref.~\cite{Giacalone:2019pca}, e.g., about the relation between initial and final anisotropies in hydrodynamics, are investigated by means of state-of-the-art hydrodynamic simulations, presented in Appendix~\ref{app:A}.

\section{Average transverse momentum and its fluctuations}
\label{sec:2}

Two nuclei colliding at relativistic energy produce a quark-gluon plasma, the hot state of strong-interaction matter. Despite being as large as an atomic nucleus, the dynamics of this entity is driven by macroscopic physics, i.e., pressure gradient forces and velocity fields. Due to an immense Lorentz contraction of the colliding nuclei in the lab frame, the created medium can be considered as invariant under longitudinal boosts~\cite{Bjorken:1982qr}, so that for most practical purposes it is enough to look at the dynamics of the system in a rapidity slice around the interaction point (midrapidity).

When the plasma decouples to particles, a spectrum of hadrons is produced at midrapidity, $\frac{dN}{d^2{\bf p}_t}$, where ${\bf p}_t$ is the particle momentum in the plane transverse to the collision axis, or \textit{transverse} plane. The quantity I shall focus on is the average transverse momentum of the hadrons emitted in a collision event:\footnote{In an actual experiment, the average transverse momentum comes from a discrete sum of a finite number of terms (particles), as discussed in Sec.~\ref{sec:33}.}
\begin{equation}
\label{eq:barpt}
\bra p_t \ket = \frac{1}{N} \int_{{\bf p}_t} \frac{dN}{d^2{\bf p}_t} p_t,
\end{equation}
where $p_t \equiv |{\bf p}_t|$, and $N=\int_{{\bf p}_t} dN/d^2{\bf p}_t$ is the total number of particles (or \textit{multiplicity}) detected in the event. The average transverse momentum can be evaluated on an event-by-event basis, and its probability distribution can be measured in experiments. 

This preliminary section is devoted to elucidating the physical origin of $\bra p_t \ket$, and of its fluctuations in nucleus-nucleus collisions. This knowledge will be crucial in the subsequent discussion of the phenomenology of nuclear structure in heavy-ion collisions.

\subsection{Relation between system size and $\bra p_t \ket$}
\label{sec:21}

Consider a uniform ideal gas of massless particles with Boltzmann statistics. The energy of a particle, $E$, in such a system is equal to $3T$, where $T$ is the temperature, while the total entropy, $S$, is proportional to the number of particles~\cite{Ollitrault:2007du}. If the system is relativistic, the energy of a particle coincides with its momentum, $p$, and one has
\begin{equation}
\label{eq:0}
    p \simeq E =  3T.
\end{equation} 
The quark-gluon plasma created in ultrarelativistic heavy-ion collisions is essentially a relativistic ideal gas of massless particles, and the existence of similar relations in the context of heavy-ion collisions, where one replaces $p$ with the final $\bra p_t \ket$, has been recently established by means of modern hydrodynamic simulations~\cite{Gardim:2019xjs}:
\begin{equation}
\label{eq:1}
    \bra p_t \ket \simeq 3 T,
\end{equation}
where $T$ is an effective temperature of the system.
\begin{figure}[t]
    \centering
    \includegraphics[width=.9\linewidth]{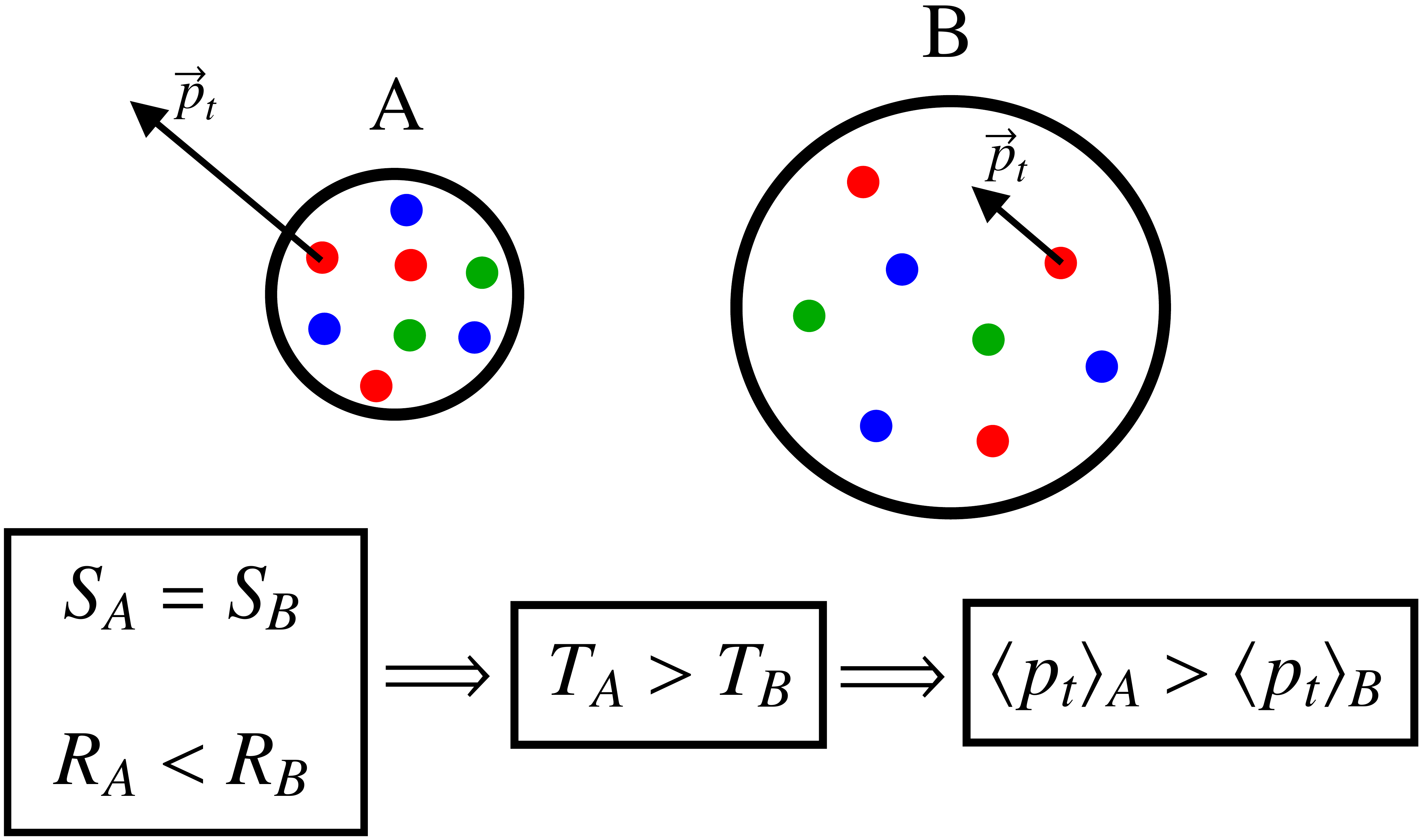}
    \caption{Relation between system size, $R$, and average transverse momentum, $\bra p_t \ket$, in a system with given entropy, $S$. Systems A and B have the same total entropy, but B has a larger volume. The temperature in A is therefore larger. Assuming an ideal classical gas of massless particles, this implies that the particles in A carry more momentum, $p_t$.}
    \label{fig:1}
\end{figure}

 Suppose one is given two uniform quark-gluon fluids that have the same entropy, but contained within different volumes. The system with a smaller volume is denser, it has larger temperature, and thus, according to Eq.~(\ref{eq:1}), it yields larger $\bra p_t \ket$ in the final state. This discussion is summarized in Fig.~\ref{fig:1}, for two systems with the same entropy but different volumes. For uniform systems with fixed number of particles, then, there is a one-to-one correspondence between the system size, which I dub $R$, and $\bra p_t \ket$, and in fact all other thermodynamic quantities (such as the energy). It is natural, then, that the origin of the fluctuations of $\bra p_t \ket$ in the literature was identified with the fluctuations of $R$~\cite{Broniowski:2009fm,Bozek:2012fw,Mazeliauskas:2015efa,Bozek:2017elk}. 

In the limit of small fluctuations, one can write an explicit relation between the fluctuations of these quantities. First write the following thermodynamic identity~\cite{Gardim:2019brr}:
\begin{equation}
   c_s^2 = \frac{dP}{d\epsilon} = \frac{d \ln T }{d \ln s}, 
\end{equation}
where $P$, $\epsilon$, $T$, and $s$ are, respectively, the pressure, energy density, temperature, and entropy density of the fluid. Then, dimensional analysis imply that $s \propto R^{-3}$, while $\bra p_t \ket \propto T$ from Eq.~(\ref{eq:1}), so that:
\begin{equation}
\label{eq:ptR}
    \frac{d \bra p_t \ket}{ \bbra p_t \kket } = -3c_s^2 \frac{d R}{\bra R \ket},
\end{equation}
where the minus sign comes from the fact that $\bra p_t \ket$ is anticorrelated with $R$. In natural units ($c=1$), the factor $3 c_s^2$ is typically of order unity. This is the relation employed in Ref.~\cite{Giacalone:2019pca} to transform the event-by-event distribution of $R$ into an event-by-event distribution of $\bra p_t \ket$. However, this picture is modified when the system is not uniform. The first important result of the present paper, described in the next subsection, is an improvement of Eq.~(\ref{eq:ptR}) via the definition of a better initial-state predictor for $\bra p_t \ket$.

\subsection{Initial energy as a predictor for $\bra p_t \ket$}
\label{sec:22}

The density profile of the quark-gluon plasma formed in relativistic nuclear collisions is neither uniform nor smooth. It is a rough landscape, that fluctuates significantly on an event-by-event basis. In presence of density fluctuations, Eq.~(\ref{eq:0}) remains valid, and $\bra p_t \ket$ remains a measure of the temperature, or of the energy, of the system; But the system size, $R$, is no longer in a one-to-one correspondence with the other thermodynamic quantities, and can not be considered as a natural predictor of $\bra p_t \ket$. 
\begin{figure}[t]
    \centering
    \includegraphics[width=.95\linewidth]{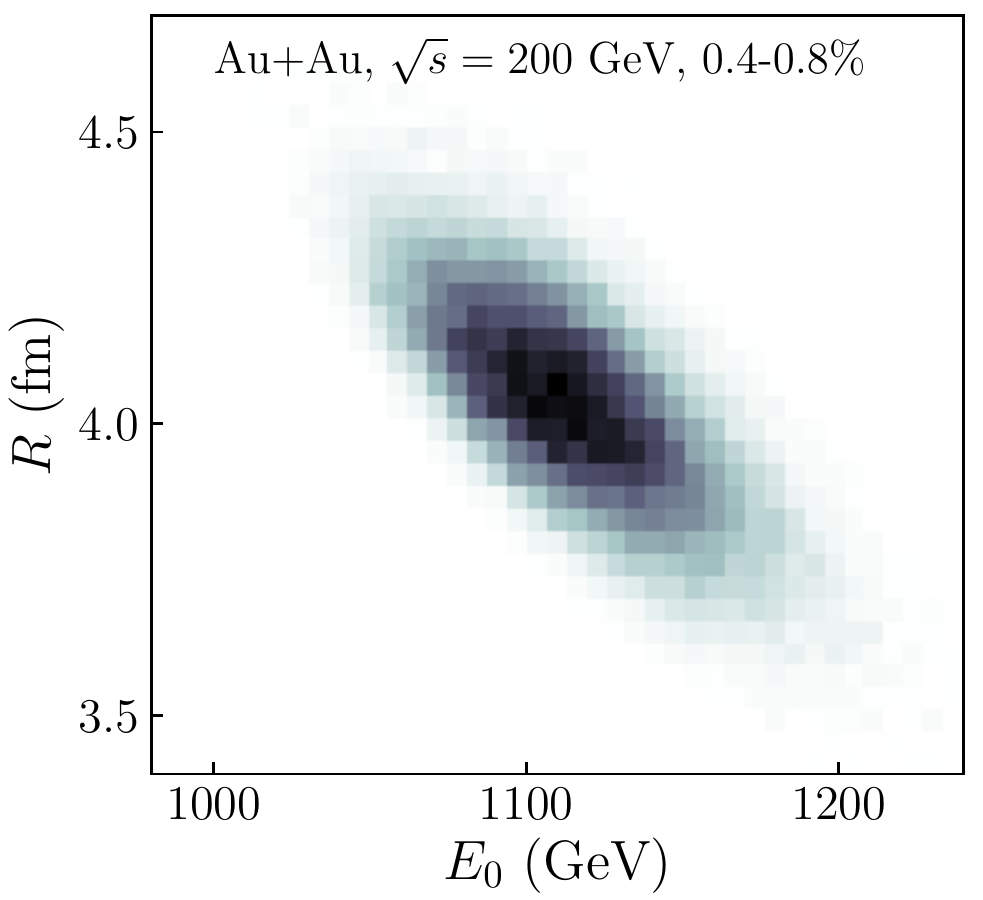}
    \caption{Joint distribution of initial energy, $E_0$, and system size, $R$, in ultracentral Au-Au collisions at top RHIC energy simulated with the \trento{} model (described in Appendix~\ref{app:tr}). Darker pixels correspond to larger numbers of counts.}
    \label{fig:2}
\end{figure}

I give an explicit illustration of this phenomenon. I define the size of the quark-gluon plasma as follows:
\begin{equation}
    R^2 = \frac{\int_{\bf x} |{\bf x}|^2 s({\bf x},\tau_0) }{\int_{\bf x} s({\bf x},\tau_0) },
\end{equation}
where ${\bf x}$ is a coordinate in the transverse plane, and $s({\bf x},\tau_0)$ is the transverse profile of entropy density at the time $\tau_0$ when hydrodynamics becomes applicable.  I also define the initial energy of the medium by:
\begin{equation}
  \label{eq:ei}
E_0 \equiv \tau_0\int_{\bf x} e({\bf x}, \tau_0),
\end{equation}
where $e({\bf x}, \tau_0)$ is the initial energy-density profile, obtained from $s({\bf x},\tau_0)$ via an equation of state. In Refs.~\cite{Gardim:2020sma,Giacalone:2020dln}, it was indeed shown that, at fixed \textit{total} entropy (per unit rapidity), $S = \tau_0 \int_{\bf x} s({\bf x},\tau_0)$, the value of $\bra p_t \ket$ at the end of hydrodynamics is in an almost perfect one-to-one correspondence with the value of $E_0$. This result was obtained in ideal hydrodynamic simulations. Here, in Appendix~\ref{app:A}, in Fig.~\ref{fig:13}, I check explicitly that this is true as well in viscous hydrodynamics. The correlation between $\bra p_t \ket$ and $R$ is instead much weaker. The reason is precisely that there is a significant dispersion between $R$ and $E_0$. 

I show this dispersion in simulations of Au+Au collisions at top RHIC energy, corresponding to nucleon-nucleon interactions at a center-of-mass energy $\sqrt{s}=200$~GeV. I do this by means of the \trento{} model of initial conditions, a parametrization of the initial entropy density, $s({\bf x},\tau_0)$, whose working principles and parameters are reported in Appendix~\ref{app:tr}. The entropy density returned by the \trento{} calculation is converted into an energy density by means, for simplicity, of the equation of state of conformal high-temperature QCD ($c_s^2(T)=1/3$):
\begin{equation}
   e({\bf x}, \tau_0)= s({\bf x}, \tau_0)^{4/3} \left ( \frac{3}{4} \right )^{4/3} \left( \nu_{\rm QCD} \frac{\pi^2}{30} \right)^{-1/3},
\end{equation}
with a number of degrees of freedom $\nu_{\rm QCD}=40$. $E_0$ is obtained upon integration of the previous equation following Eq.~(\ref{eq:ei}). Figure~\ref{fig:2} shows the joint distribution of $E_0$ and $R$ in a sample of \textit{ultracentral} Au+Au collisions, as defined in Appendix~\ref{app:tr}, where the value of $S$ is essentially fixed. The energy and the size are clearly negatively correlated, but they are not in a one-to-one correspondence. Therefore, if $E_0$ is the same thing as $\bra p_t \ket$, then $\bra p_t \ket$ is not the same thing as $R$.

In this paper, I use $E_0$ as a predictor for $\bra p_t \ket$ to improve Eq.~(\ref{eq:ptR}), and thus the results of Ref.~\cite{Giacalone:2019pca}. Note that in Fig.~\ref{fig:2}, and in general throughout this manuscript, the value of $S$ is not strictly fixed, but the considered events belong to a narrow interval of $S$ (as explained in detail in Appendix~\ref{app:tr}). This is a better representation of the \textit{centrality} selection performed in experiments. To take the small fluctuations of $S$ into account, one can simply replace $E_0$ with $E_0/S$.\footnote{Note that this choice is not unique. The authors of Ref.~\cite{Schenke:2020uqq} show, for instance, that $E_0/A$, where $A$ is the area of the system, works just as well, or better, than $E_0/S$.} The improved Eq.~(\ref{eq:ptR}) is finally given by:
\begin{equation}
\label{eq:predictor}
   \frac{d\bra p_t \ket}{\bbra p_t \kket} =  \kappa_0 \frac{d(E_0/S)}{\bra E_0/S \ket}  ,
\end{equation}
where $\kappa_0$ is a phenomenological parameter, which can be inferred from experimental data on the relative fluctuation of $\bra p_t \ket$. The latter was determined recently by the STAR collaboration~\cite{Adam:2019rsf}. They measured the relative fluctuation of $\bra p_t \ket$ originating from genuine \textit{collective} effects, such as those predicted by the \trento{} model. They found that the left-hand side of Eq.~(\ref{eq:predictor}) is equal to:
\begin{equation}
\label{eq:sigmadyn}
    \frac { \sigma_{\rm dynamical}(\bra p_t \ket)}{\bbra p_t \kket} = 0.012,
\end{equation}
in central collisions at top RHIC energy. The relative fluctuation of $E_0/S$, i.e., the right-hand side of Eq.~(\ref{eq:predictor}), in my \trento{} calculation is instead of order 0.03, so that in practice one has $\kappa_0\approx0.4$.

The conclusion of this section is that a natural predictor of $\bra p_t \ket$ in presence of initial density fluctuations is provided by the initial energy, and not by the initial size. Throughout this manuscript, I assume that the distribution of $\bra p_t \ket / \bbra p_t \kket - 1 $ at a given centrality is equal to the distribution of $(E_0/S)/ \bra E_0/S \ket - 1 $, multiplied by a factor $\kappa_0$ to match the width of the dynamical $\bra p_t \ket$ distribution observed in data.

\section{Nuclear deformation in heavy-ion collisions}
\label{sec:3}
\begin{figure}[t]
    \centering
    \includegraphics[width=.65\linewidth]{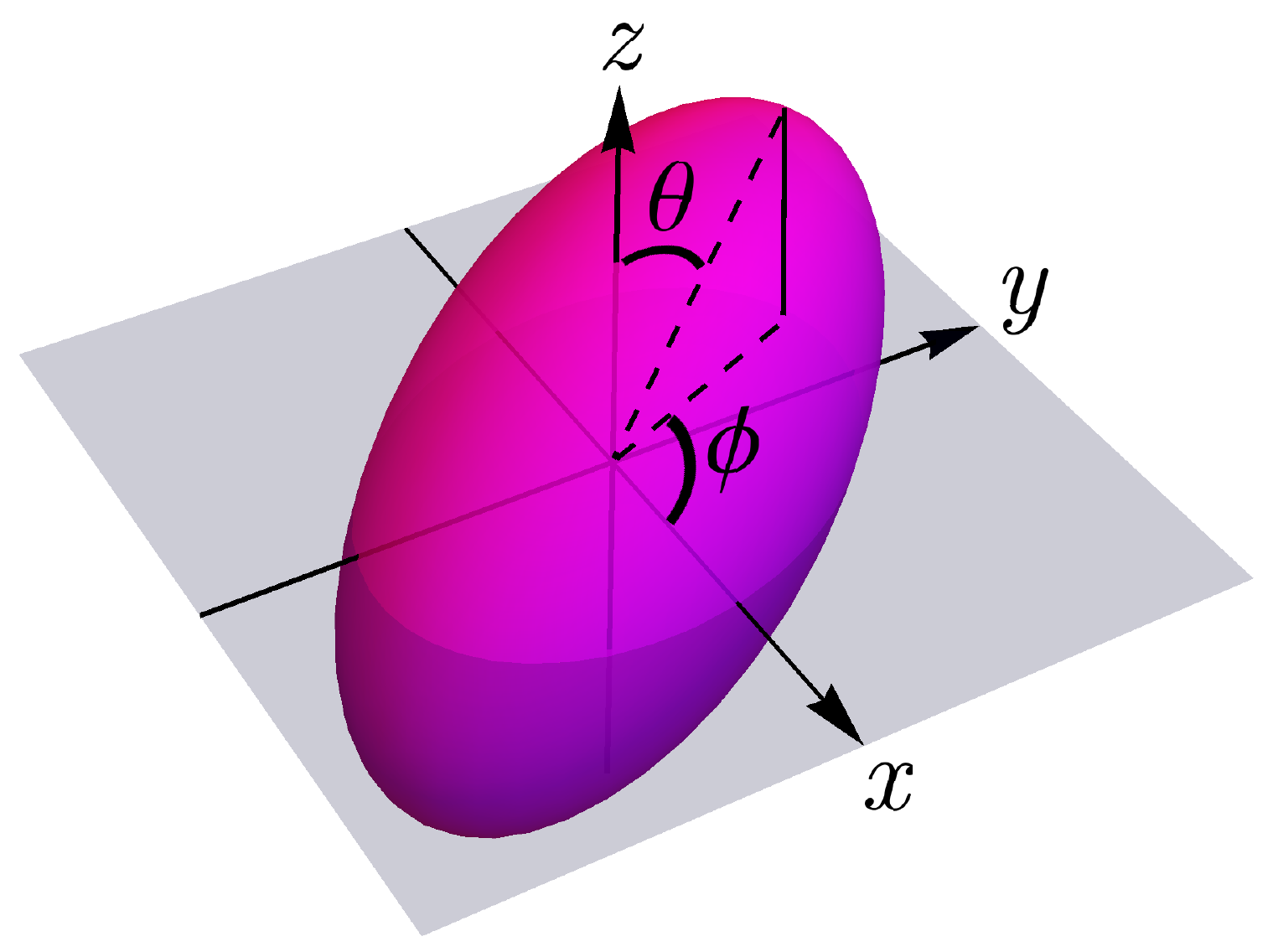}
    \caption{A deformed nucleus with $\beta>0$ randomly oriented in the laboratory frame. With respect to the intrinsic frame of the nucleus, the ellipsoid is rotated by a polar tilt, $\theta$, and by an azimuthal spin, $\phi$. In a heavy-ion collision experiment, $z$ represents the beam axis, while $(x,y)$ is the transverse plane.}
    \label{fig:3}
\end{figure}

The goal of this section is to reproduce the results of Ref.~\cite{Giacalone:2019pca} with improved theoretical ingredients, in particular, implementing the predictor for the fluctuations of $\bra p_t \ket$ introduced in the previous section. I establish a phenomenology of nuclear structure in relativistic heavy-ion collisions centered around the fact that $\bra p_t \ket$ can be used as a tool to access the shape of the colliding nuclear species. This provides the platform for the quantitative results and model-to-data comparisons shown in Sec.~\ref{sec:4}.

\subsection{Modeling the colliding bodies}

The modeling of the structure of the colliding ions is simple, and follows typically the Bohr-Mottelson model~\cite{BM}. The average mass density in the nucleus is defined by:
\begin{equation}
\label{eq:density}
    \rho({\bf x}',z') = \frac{\rho_0}{1+\exp \biggl\{ \frac{1}{a} \biggl[  \sqrt{|\textbf{x}'|^2+z'^2} - R \bigl(1 + \beta Y_{2}^0 \bigr)  \biggr] \biggr\} },
\end{equation}
where $z'$ is the direction of the nuclear axis, ${\bf x}'$ is the plane orthogonal to it. Spherical symmetry is broken by the spherical harmonic $Y_2^0$. $R$ is the nuclear radius, $a$ is the diffusiveness, and $\beta$ is the quadrupole deformation parameter,\footnote{Note that it is more customary to truncate the multipole expansion at the hexadecapole deformation, $Y_4^0$, which introduces an additional deformation parameter, $\beta_4$. Here I neglect the effects of this deformation, which should be unimportant in studies of ultracentral collisions~\cite{Shou:2014eya}.}. In particular, one has~\cite{Kumar:1972zza,Schmidt:2017afd}: 
\begin{equation}
    \beta \simeq \frac{4\pi}{5} \frac{\int d^3{\bf r}~Y_2^0 (\Theta, \Phi) r^2 \rho({\bf r})}{\int d^3{\bf r}~ r^2 \rho({\bf r})},
\end{equation}
where $\rho({\bf r})$ corresponds to Eq.~(\ref{eq:density}). If $\beta>0$, the nucleus is \textit{prolate}, while it is \textit{oblate} for $\beta<0$. 
\begin{figure*}[t]
    \centering
    \includegraphics[width=.7\linewidth]{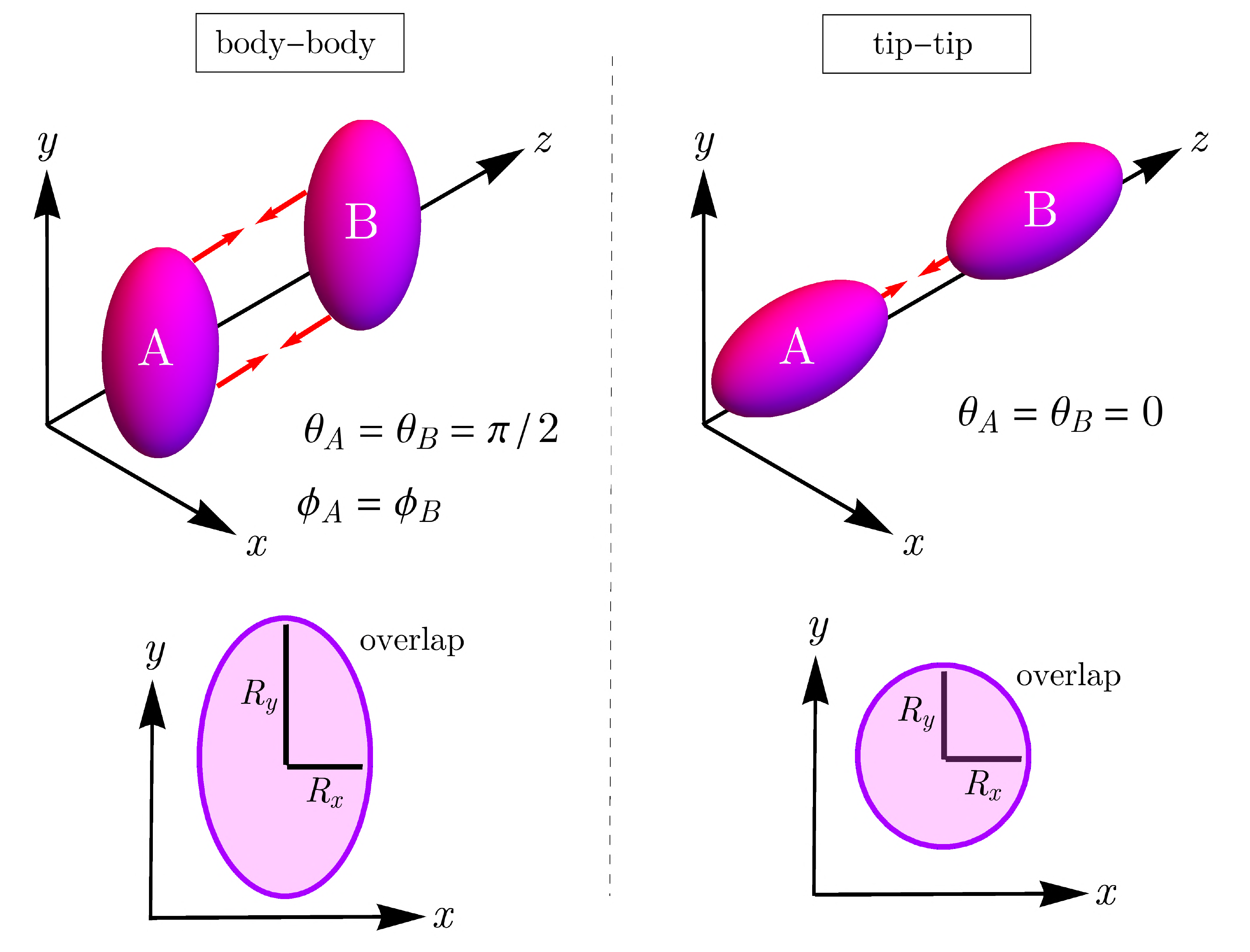}
    \caption{Left: illustration of a fully-overlapping body-body collision (top). The nuclei collide along the beam axis, $z$, with their axes aligned orthogonally to $z$ ($\theta_A=\theta_B=\pi/2$), and with the same azimuthal orientation, $\phi$. The transverse area of nuclear overlap in a body-body collision (bottom) possesses an enhanced quadrupolar asymmetry ($R_y > R_x$), as it follows closely the deformed shape of the colliding bodies. Right: illustration of a fully-overlapping tip-tip collision (top). The nuclei collide with $\theta_A=\theta_B=0$.  The area of overlap (bottom) has the symmetry of a disk, $R_x = R_y$.}
    \label{fig:4}
\end{figure*}

The fluctuations of the mass density are simply obtained by a Monte Carlo sampling of nucleon coordinates~\cite{Miller:2007ri}, distributed according to Eq.~(\ref{eq:density}), performed on an event-by-event basis. Note that the quadrupole deformation is static, in the sense that the sampling of nucleons is performed from an average density which has always the same value of $\beta$. Doing so one neglects the fact that the $\beta$ can have a significant degree of softness~\cite{Mustonen:2018ody,Poves:2019byh}. In this paper, I will be concerned mostly with $^{238}$U nuclei, for which this is a good approximation~\cite{Poves:2019byh}, but it would be interesting to include fluctuating values of $\beta$, as well as of a potential triaxial deformation parameter, in collisions of smaller nuclei, such as $^{96}$Ru, $^{96}$Zr, or $^{129}$Xe, for which experimental data is available.

The deformed nucleus is now injected in the beam pipe of a particle accelerator. The orientation of the ellipsoid is random, so that in general the intrinsic nuclear frame $(z', {\bf x'})$ differ from the laboratory frame $(z,{\bf x})$ by a polar tilt, $\theta$, and by an azimuthal spin, $\phi$. This is illustrated in Fig.~\ref{fig:3}. Consequently, the collision geometry for two colliding nuclei, A and B, in the laboratory frame is characterized by two polar tilts, $\theta_A$ and $\theta_B$, and two azimuthal rotations, $\phi_A$ and $\phi_B$.

\subsection{$\bra p_t \ket$ and collision geometry: \\ tip-tip vs. body-body collisions}
\label{sec:31}

I explain now the argument introduced in Ref.~\cite{Giacalone:2019pca} that the average transverse momentum of hadrons produced in a heavy-ion collision can be used as a tool to somehow \textit{freeze} the orientation of the colliding nuclei. The focus is always on ultracentral collisions, i.e., on collisions where almost all the nucleons from the colliding ions participate in the interaction, and thus the overlap of the two nuclei is maximal. Two kinds of fully-overlapping collisions are particularly interesting, as illustrated in Fig.~\ref{fig:4}.
\begin{itemize}
    \item There are \textit{body-body} configurations, shown in the left panel of Fig.~\ref{fig:4}, in which the axes of the two nuclei are both orthogonal to the beam direction, i.e., $\theta_A=\theta_B=\pi/2$, and both nuclei are rotated by the same azimuthal angle, $\phi_A=\phi_B$. As shown in the Fig.~\ref{fig:4}, in such configurations the area of overlap in the transverse plane has an enhanced elliptical deformation, which originates from the shape of the colliding nuclei.
    \item  There are \textit{tip-tip} configurations, shown in the right panel of Fig.~\ref{fig:4}, in which the axis of both nuclei is aligned with the beam axis, $z'= z$, or $\theta_A=\theta_B=0$. The resulting area of overlap is circular.
\end{itemize}
\begin{figure*}[t]
    \centering
    \includegraphics[width=.95\linewidth]{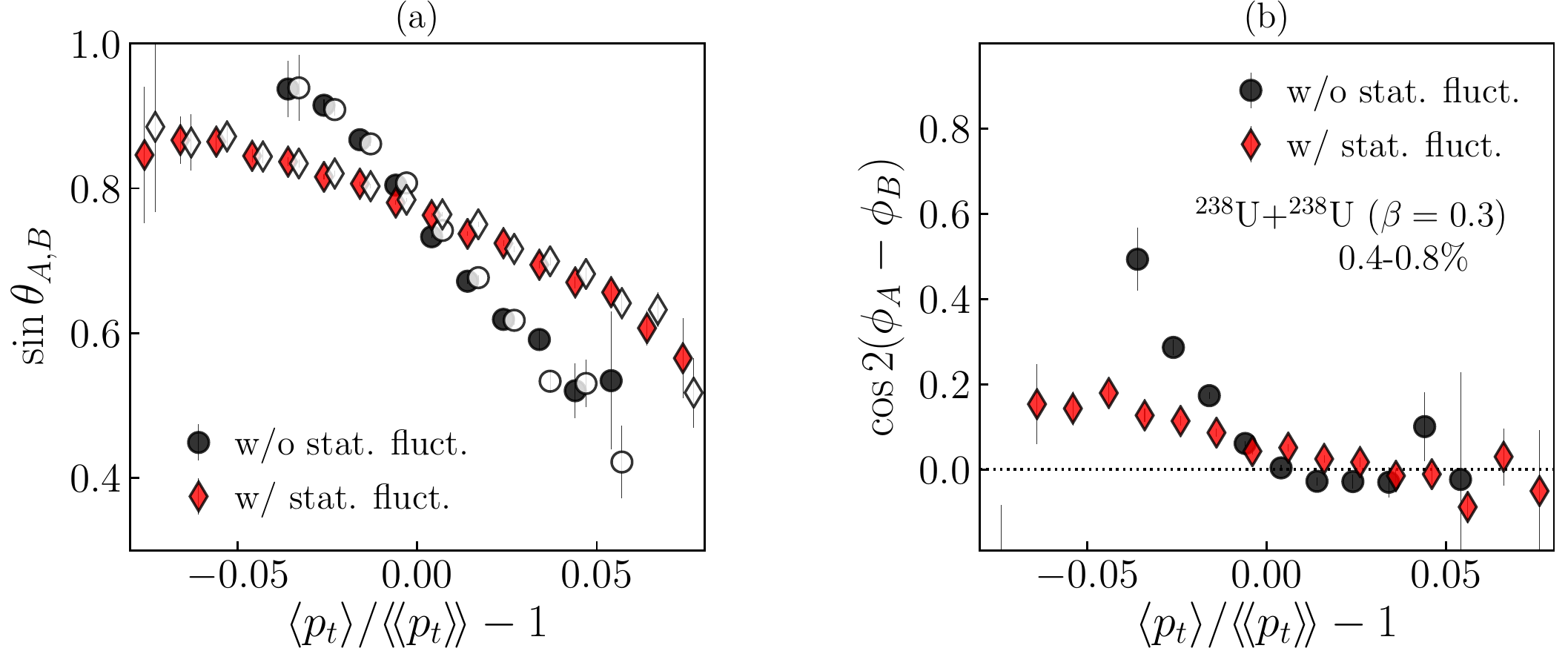}
    \caption{(a) Sine of $\theta_A$ and $\theta_B$ as function of $\bra p_t \ket$ in ultracentral collisions of deformed $^{238}$U nuclei ($\beta=0.3$) at top RHIC energy. Circles represent the results obtained without the inclusion of statistical fluctuations, as explained in Sec.~\ref{sec:33} (full circles: $\theta_A$, empty circles: $\theta_B$). Such fluctuations are instead included in the results shown as diamonds (same color coding for A and B). Symbols are shifted horizontally for readability. (b) Azimuthal alignment, $\cos2(\phi_A - \phi_B)$, of the colliding $^{238}$U nuclei. Diamonds include the effect of statistical fluctuations. See Appendix~\ref{app:tr} for details about the \trento{} implementation leading to these results.}
    \label{fig:5}
\end{figure*}

The realization of Ref.~\cite{Giacalone:2019pca} can be formulated as follows. If we look at fully-overlapping events at fixed final-state multiplicity (i.e., fixed initial entropy), then tip-tip collisions correspond to the configurations where the largest densities are achieved. Compared to body-body collisions, a tip-tip event deposits the same amount of entropy in a smaller volume. Consequently, the temperature (and the energy) of the tip-tip event is larger, and thus, if the number of particles is fixed, one naturally expects $\bra p_t \ket$ to be also larger [Eq.~(\ref{eq:1})]. The idea is that, in a given sample of events at fixed multiplicity, large values of $\bra p_t \ket$ are mainly produced by configurations that are tip-tip-like, while low values of $\bra p_t \ket$ come from body-body configurations.\footnote{ In Appendix~\ref{app:A}, I show the initial energy-density profile of a tip-tip collision and of a body-body collision as an illustration of the above-mentioned phenomenon for which tip-tip events achieve larger densities of energy (see Fig.~\ref{fig:12}). Additionally, I carry out hydrodynamic simulations to show explicitly that tip-tip collisions yield larger $\bra p_t \ket$ than body-body collisions (see Fig.~\ref{fig:13}).}

Following Ref.~\cite{Giacalone:2019pca}, I provide now an explicit confirmation that this idea works in the \trento{} model. I collide $^{238}$U nuclei at top RHIC energy. In Ref.~\cite{Raman:1201zz}, it is reported that these nuclei have a significant prolate deformation, $\beta=0.286$, which I round up to 0.3 for simplicity. I focus on ultracentral collisions, as described in Appendix~\ref{app:tr}. I compute the distribution of $(E_0/S)/\bra E_0/S \ket -1 $, in my sample, and I turn it into the distribution of $\bra p_t \ket/\bbra p_t \kket -1 $ by means of Eq.~(\ref{eq:predictor}).

In Fig.~\ref{fig:5}(a), I show the sine of the polar tilts, $\sin \theta_{A,B}$, as a function of the relative variation of $\bra p_t \ket$. The results are shown as circles. We see that the value of the sine quickly grows to unity as soon as we move towards low values of $\bra p_t \ket$, which confirms our expectation that low $\bra p_t \ket$ selects body-body configurations, i.e., $\theta_A=\theta_B=\pi/2$. Note that $\sin \theta$ does not go to zero at the largest values of $\bra p_t \ket$, meaning that, on average, tip-tip collisions are not achieved. 

In Fig~\ref{fig:5}(b), I look instead at the alignment of the colliding nuclei in the azimuthal plane, quantified by the correlation $\cos 2(\phi_A-\phi_B)$. This result is shown as circles. Moving towards the low-$\bra p_t \ket$ tail, I observe that the correlation becomes positive, reaching a value of 0.5. This implies that a selection of fully-overlapping body-body collisions is effectively taking place at low $\bra p_t \ket$.

I conclude that one can indeed have an handle on the orientation of the colliding bodies by sorting ultracentral events according to their value of $\bra p_t \ket$. Note that the curves in both Fig.~\ref{fig:5}(a) and Fig.~\ref{fig:5}(b) are significantly flatter than those shown in Fig.~3 of Ref.~\cite{Giacalone:2019pca}. This is due to the fact that I am now consistently taking into account the fact that $R$ and $\bra p_t \ket$ are not in a one-to-one correspondence.

\subsection{Anticorrelation between $\varepsilon_2$ and $\bra p_t \ket$}
\label{sec:32}

The second point made in Ref.~\cite{Giacalone:2019pca} is that the capability of discerning between body-body and quasi-tip-tip collisions has a very important phenomenological consequence.  As illustrated in Fig.~\ref{fig:4}, body-body collisions have a different \textit{geometry} than tip-tip collisions, as they are characterized by a elliptical asymmetry, depending on the value of $\beta$. 
 Let me introduce the initial \textit{eccentricity} of the quark-gluon plasma, defined as the second Fourier harmonic of the entropy density profile~\cite{Teaney:2010vd}: 
 \begin{equation}
 \label{eq:eps2}
     \varepsilon_2 = \frac{\left | \int_{\bf x} {\bf x}^2 s({\bf x},\tau_0) \right |}{\int_{\bf x} |{\bf x}|^2 s({\bf x},\tau_0)},
 \end{equation}
where ${\bf x}^2$ in the numerator should be read in complex notation, ${\bf x}^2=(x+ iy)^2$.
\begin{figure}[t]
    \centering
    \includegraphics[width=.95\linewidth]{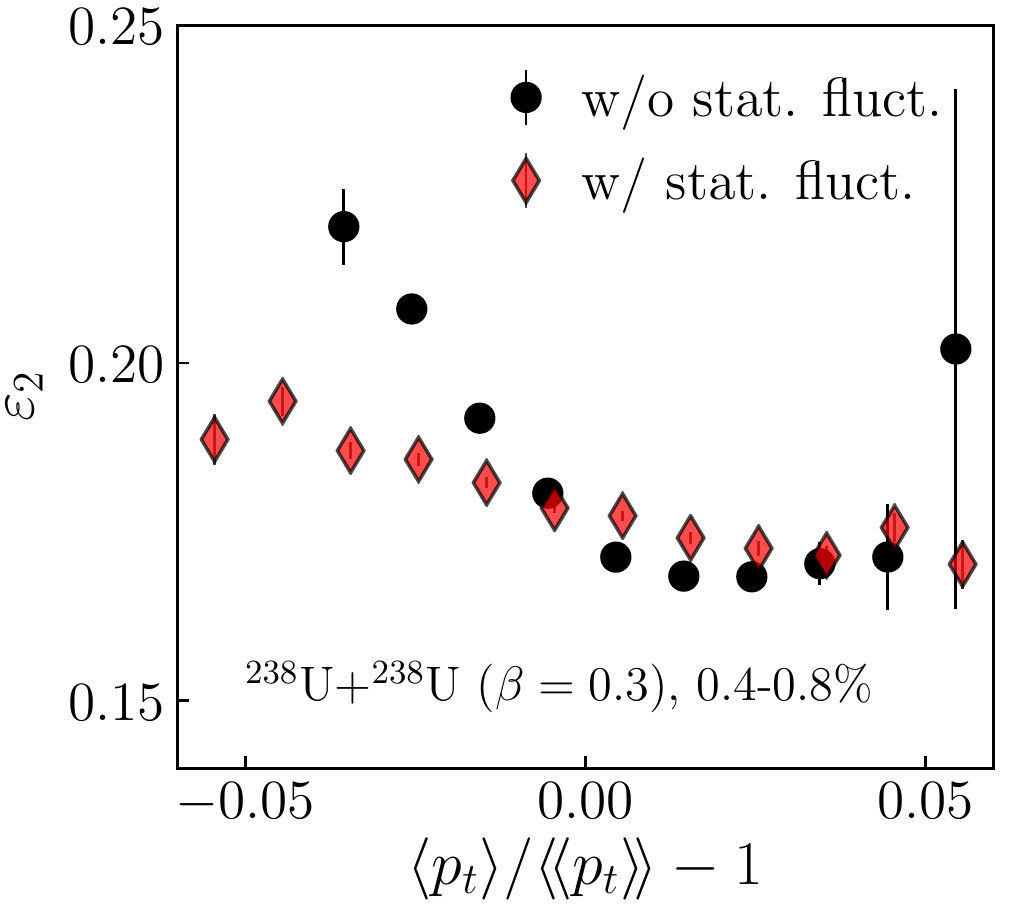}
    \caption{Rms eccentricity as function of $\bra p_t \ket$ in ultracentral U+U collisions at top RHIC energy. Diamonds include the effect of statistical fluctuations explained in Sec.~\ref{sec:33}.}
    \label{fig:6}
\end{figure}

The results of Fig.~\ref{fig:5}, combined with the illustration in Fig.~\ref{fig:4}, imply that collisions at low $\bra p_t \ket$, which probe body-body geometries, should have a larger $\varepsilon_2$ than collisions at high $\bra p_t \ket$. Using my sample of ultracentral U+U collisions from the \trento{} calculation, I plot the rms value of $\varepsilon_2$ as a function of $\bra p_t \ket$ in Fig.~\ref{fig:6}. The result is shown as black circles. One sees that $\varepsilon_2$ decreases quickly as a function of $\bra p_t \ket$, confirming the intuitive picture.  

The result in Fig.~\ref{fig:6} is one of the most important in this paper, because one can easily relate the initial spatial anisotropy, $\varepsilon_2$, to the measured final-state anisotropy in momentum space, $v_2$, to be defined in Sec.~\ref{sec:4}, and thus obtain a result expressed solely in terms of measurable observables. That is eventually my goal, but before doing so, an additional ingredient needs to be included in the present analysis.

\subsection{Effect of statistical fluctuations}
\label{sec:33}

 Evaluating quantities as function of $\bra p_t \ket$ requires the knowledge of the mean transverse momentum in each event. While this is not an issue in a hydrodynamic simulation, where the output of one event is a continuous spectrum in momentum space, the situation is different in an experiment, where the equivalent of Eq.~(\ref{eq:barpt}) is given by a discrete average:
\begin{equation}
    \bra p_t \ket = \frac{1}{N}\sum_{i=1}^{N} p_{t,i},
\end{equation}
where $N$ is the number of particles detected in the event, and $p_{t,i}$ is the transverse momentum of the $i$th particle. The average momentum is evaluated from the sum of a finite number of terms, typically of order 1000 in central U+U collisions, and its determination is therefore affected by a statistical error. This implies in particular that, in an experiment, the event-by-event correspondence between $\bra p_t \ket$ and $E_0$ is smeared by the finite number of detected particles. Statistical fluctuations are naturally proportional to $1/\sqrt{N}$, and, relative to the mean value of $N$, are typically of order 1\% for $N\approx 1000$. However, this is as large as the relative dynamical fluctuation of $\bra p_t \ket$, which is about 1.2\% in central collisions at RHIC~\cite{Adam:2019rsf} [see Eq.~(\ref{eq:sigmadyn})].

The magnitude of statistical fluctuations can be evaluated from the law of large numbers. In a heavy-ion collisions, particles are emitted independently from the decoupling surface, with a random value of transverse momentum, $p_t$, taken from an underlying probability distribution, i.e., the $p_t$ spectrum, which can be measured at a given collision centrality. Therefore, the fluctuation of the average transverse momentum due to the finite number of particles is given by:
\begin{equation}
    \sigma_{\rm stat}=\frac{1}{\sqrt{N}}\sqrt{\bra p_t^2 \ket - \bra p_t \ket^2},
\end{equation}
where the average is weighted with respect to the measured spectrum, $\bra \ldots \ket = \frac{1}{N} \left [ \int \ldots \frac{dN}{d^2{\bf p}_t} \right ]$. Considering that at top RHIC energy in the full acceptance of the STAR detector, $|\eta|<1$, one detects $N \approx 1000$ particles, and using the spectrum measured in central Au+Au collisions~\cite{Back:2003qr}, I obtain:
\begin{equation}
    \sigma_{\rm stat} = 0.01~{\rm GeV}.
\end{equation}
The relative statistical fluctuation is, hence:
\begin{equation}
\label{eq:statrel}
    \sigma_{\rm stat}/\bbra p_t \kket = 0.18,
 \end{equation}
where I used $\bbra p_t \kket=0.57$~GeV~\cite{Adams:2005ka}. This is the relative fluctuation of $\bra p_t \ket$ coming from the simple fact that $N$ is not infinite.

This result has to be compared with the magnitude of the relative dynamical fluctuation of $\bra p_t \ket$. Comparing Eq.~(\ref{eq:statrel}) with Eq.~(\ref{eq:sigmadyn}), one finds:
\begin{equation}
\label{eq:dynstat}
    \frac{\sigma_{\rm stat} (\bra p_t \ket)}{\bbra p_t \kket} = 1.5 \times \frac{\sigma_{\rm dynamical} (\bra p_t \ket)}{\bbra p_t \kket}.
\end{equation}
Relative statistical fluctuations are, hence, somewhat larger than relative dynamical fluctuations. Any theoretical prediction requiring the evaluation of $\bra p_t \ket$ on an event-by-event basis is meaningless unless it takes this statistical smearing into account.

Including the effect of these statistical fluctuations in a theoretical calculation is in fact trivial. Since particles are emitted independently at freezeout, their number follows a Poisson distribution, or equivalently, a Gaussian distribution, because $N \gg 1$. Therefore, I can readily correct my \trento{} results for statistical fluctuations. First, I use Eq.~(\ref{eq:predictor}) to compute the distribution of $\bra p_t \ket/\bbra p_t \kket$, whose width, say $\sigma$, is given in Eq.~(\ref{eq:sigmadyn}). Then, I multiply each entry of this distribution by a random number drawn from a Gaussian distribution of unit mean and width given by $1.5 \times \sigma$, following Eq.~(\ref{eq:dynstat}). This produces a distribution of fictitious values of $\bra p_t \ket/\bbra p_t \kket$, and if I now evaluate the final observables as function of this corrected quantity, the decorrelation between $E_0/S$ and $\bra p_t \ket$ due the finite-$N$ effect will be properly included.

I implement this correction, and I show its impact on the observables analyzed in this section. In Fig.~\ref{fig:4}(a), the new results are shown as diamonds. The effect of the statistical smearing is very visible. As expected, it washes out much of the correlation between $\sin \theta_{A,B}$ and $\bra p_t \ket$, yielding a curve with a smaller slope. An even larger effect is observed in Fig.~\ref{fig:4}(b) (diamonds), where the alignment between azimuthal angles is almost entirely washed out by the statistical fluctuations. The slope of $\varepsilon_n$ as a function of $\bra p_t \ket$, now shown as diamonds in Fig.~\ref{fig:6}, is also significantly reduced. Note that statistical fluctuations make the $\bra p_t \ket$ distribution broader.

However, despite this large correction, the conclusion of Ref.~\cite{Giacalone:2019pca} remains valid. The \trento{} model clearly predicts that $\bra p_t \ket$ is anticorrelated with $\varepsilon_2$ in ultracentral collisions of $^{238}$U nuclei with $\beta=0.3$. Note that one can also define a measure of the correlation between anisotropy and $\bra p_t \ket$ which is by construction insensitive to the statistical fluctuations analyzed here. This will be discussed in Sec.~\ref{sec:42}. 

Let me move on, then, to a quantitative phenomenological study.

\section{Correlation between $\boldsymbol{\bra p_t \ket}$ and $\boldsymbol{v_n}$: \\ quantitative analysis}
\label{sec:4}

To exhibit results that can be compared to experimental data, I need to convert $\varepsilon_2$ into a quantity that can be measured. This can be done by means of the \textit{elliptic flow} of the final-state spectrum. In polar coordinates,
\begin{equation}
    \frac{dN}{d^2{\bf p}_t} = \frac{dN}{p_t d p_t d \phi_p},
\end{equation}
and elliptic flow is defined as the second (complex) harmonics of the angular part:
\begin{equation}
\label{eq:flow}
    V_2 = \frac{1}{N} \int_{{\bf p}_t} \frac{1}{2\pi} \frac{dN}{d^2{\bf p}_t} e^{-i 2 \phi_p}.
\end{equation}

Elliptic flow originates from a quadrupole-like imbalance of pressure gradient forces, $\vec F=- \vec \nabla P$, within the quark-gluon medium~\cite{Ollitrault:1992bk}. This occurs when the spatial distribution of deposited matter has an elliptical asymmetry. Following Fig.~\ref{fig:4}, consider a medium that is elongated along a certain transverse direction, e.g., $R_y>R_x$. Pressure gradient forces in the fluid scale with the inverse of the length, so that one has:
\begin{equation}
    R_x < R_y \Rightarrow F_x > F_y.
\end{equation}
Under this condition, then, the hydrodynamic flow builds more momentum along $x$ than along $y$, so that the particles emitted at the end of hydrodynamics have more momentum along $x$ than along the direction orthogonal to it. This phenomenon defines elliptic flow. Hence, body-body collisions in Fig.~\ref{fig:4} naturally produce a larger elliptic flow than tip-tip collisions.

Hydrodynamic simulations~\cite{Niemi:2015qia} show that, much as $E_0$ can be used as a predictor of $\bra p_t \ket$, the value of $v_2 \equiv |V_2|$ in each event can be traced back to the value of $\varepsilon_2$, given by Eq.~(\ref{eq:eps2}), at the beginning of the hydrodynamic expansion. At a given collision centrality, their rms values can be related through a simple linear scaling, $ v_2 = \kappa_2 \varepsilon_2$. This provides a convenient and powerful method to estimate elliptic flow at a given centrality without running numerically-expensive hydrodynamic simulations, provided that the value of $\kappa_2$ is known. Note that linear scaling is especially valid when $\varepsilon_2$ is not large~\cite{Noronha-Hostler:2015dbi}, which is typically the case in ultracentral collisions.\footnote{This is not necessarily true in body-body collisions, as I discuss in Appendix~\ref{app:A}.}

The same procedure works as well for the third harmonic. A nonzero initial triangular deformation, $\varepsilon_3$, yields a triangular flow of quark-gluon matter, and thus a nonzero third harmonic in the final hadron spectrum, $v_3$~\cite{Alver:2010gr}. The two quantities are in a linear relation, $v_3=\kappa_3 \varepsilon_3$. Note that unlike $\varepsilon_2$, the triangular deformation is not generated by a symmetry of the problem (unless one collides head-on nuclei that have an intrinsic octupole moment), but solely from density fluctuations.
\begin{figure*}[t]
    \centering
    \includegraphics[width=.9\linewidth]{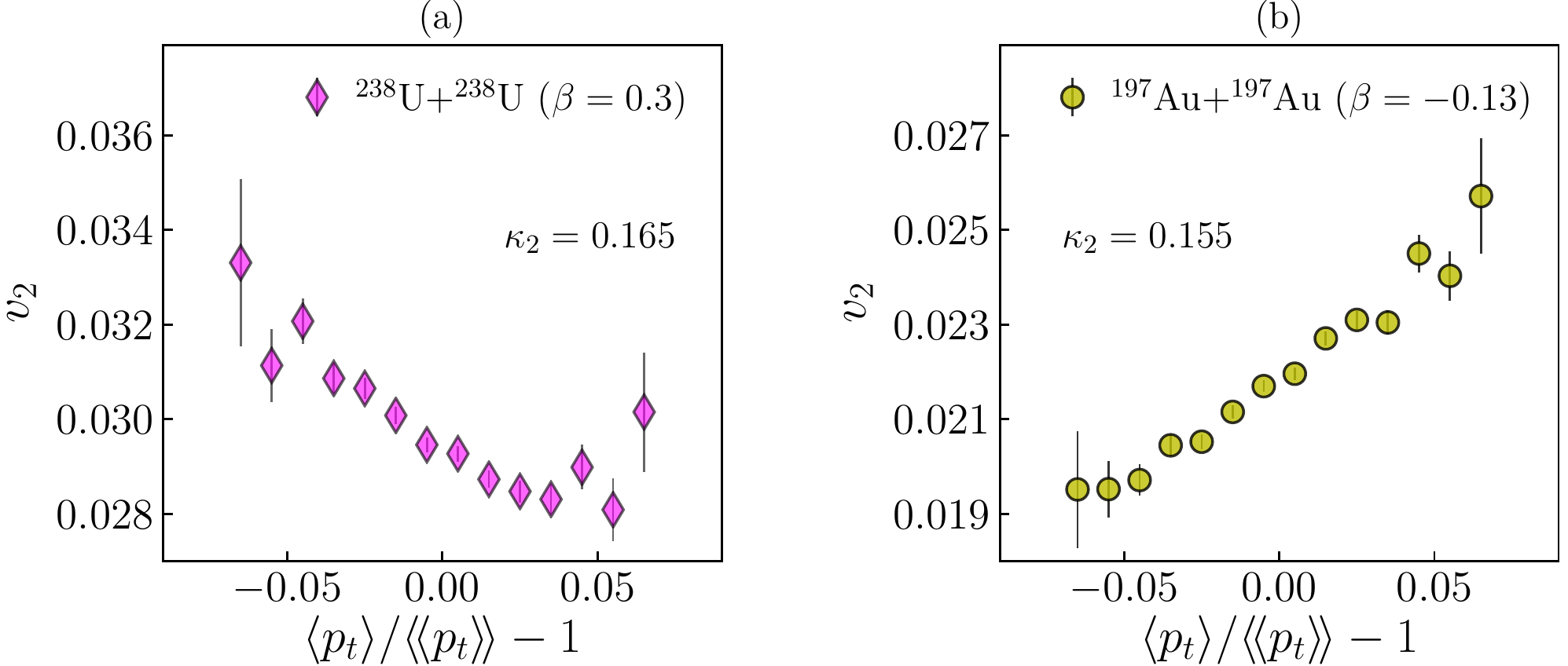}
    \caption{Rms elliptic flow coefficient, $v_2$, as a function of $\bra p_t \ket$ in $\sqrt{s}=200$~GeV U+U and Au+Au collisions. (a) $v_2$ in U+U collisions. (b) $v_2$ in Au+Au collisions. The response coefficient $\kappa_2=v_2/\varepsilon_2$, given by Eq.~(\ref{eq:kappan2}), is specified in each panel. The implementation of the \trento{} model used to obtain these results is detailed in Appendix~\ref{app:tr}.}
    \label{fig:7}
\end{figure*}

The remainder of this paper is devoted to a phenomenological study of the correlation between $\bra p_t \ket$ and $v_n$ in U+U and Au+Au collisions. I evaluate $v_2$ and $v_3$ through linear scaling,
\begin{equation}
\label{eq:kappan1}
    v_n = \kappa_n\varepsilon_n,
\end{equation}
where, depending on the context, $v_n$ indicates either the elliptic flow computed in one event or its rms value at a given $\bra p_t \ket$.  I use:
\begin{align}
\label{eq:kappan2}
\nonumber    \kappa_2[{\rm U+U}]=0.165, \hspace{20pt}\kappa_2[{\rm Au+Au}]=0.155, \\
    \kappa_3[{\rm U+U}]=0.110, \hspace{20pt}\kappa_3[\rm{ Au+Au}]=0.100.    
\end{align}
The choice of these values is motivated by the results of full hydrodynamic simulations that I carry out in Appendix~\ref{app:A}. Note that in Ref.~\cite{Giacalone:2019pca} the response coefficients were only guessed on the basis of earlier comparisons between the \trento{} model and experimental data. Here, I am explicitly checking consistency between the employed values of $\kappa_n$ and the hydrodynamic framework of nucleus-nucleus collisions.

\subsection{$v_n$ as a function of $\bra p_t \ket$}

\label{sec:41}

Figure~\ref{fig:7} shows quantitative predictions for the rms elliptic flow in ultracentral U+U and Au+Au collisions at top RHIC energy, as function of the relative variation of $\bra p_t \ket$. The prediction in Fig.~\ref{fig:7}(a) corresponds to the results of Fig.~\ref{fig:6}, rescaled by a factor $\kappa_2=0.165$, and including the effect of statistical fluctuations. Figure~\ref{fig:7}(b) shows instead predictions for Au+Au collisions. The correlation is in this case positive, and this is not surprising. The correlation between mean transverse momentum and anisotropic flow has been recently studied in collisions of spherical $^{208}$Pb nuclei at LHC energy by the ATLAS collaboration~\cite{Aad:2019fgl}. They have indeed observed that $v_2$ and $\bra p_t \ket$ are positively correlated in ultracentral collisions.

The results in Fig.\ref{fig:7} should, hence, be compared to preliminary STAR data~\cite{shengli}. One finds that my predictions provide a good description of data. The magnitude of elliptic flow is captured within an accuracy of 10\%, and the slopes of $v_2$ versus $\bra p_t \ket$ are also nicely reproduced. In particular, STAR preliminary data presents a nontrivial \textit{negative} slope in U+U collisions, confirming the overall picture of this paper, and so the idea of Ref.~\cite{Giacalone:2019pca} about the selection of body-body-like events at low $\bra p_t \ket$. I stress that my predictions do not have any free parameters. The features of the model are constrained by other sets of data, except for the value of $\beta$, which is taken from nuclear data literature. Note that this calculation implements mildly-oblate $^{197}$Au nuclei with $\beta=-0.13$. The argument explained in Sec.~\ref{sec:31} suggests that the oblate deformation yields a positive slope, enhancing the correlation between $\bra p_t \ket$ and $v_2$. In Appendix~\ref{app:B}, I assess the impact of $\beta$ on this result, by studying collisions of spherical $^{197}$Au nuclei.

\begin{figure}[b]
    \centering
    \includegraphics[width=\linewidth]{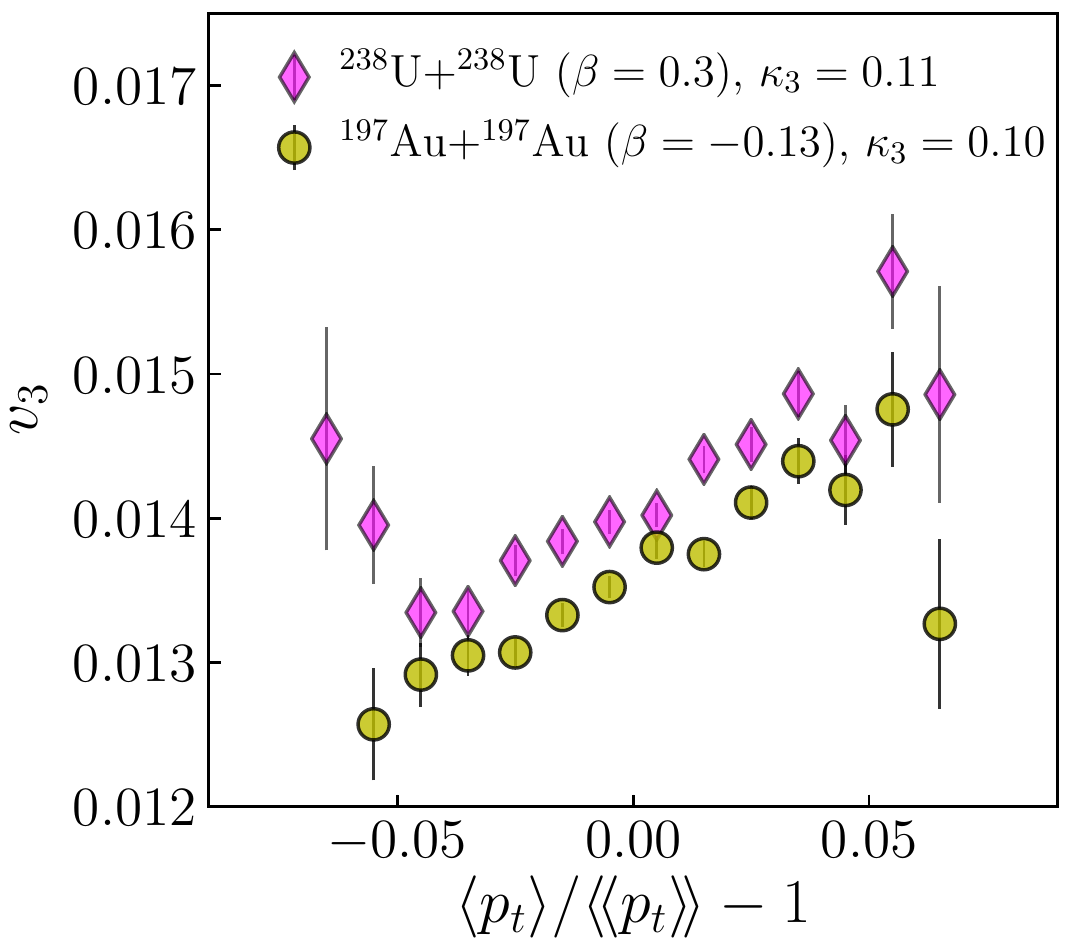}
    \caption{Predictions for the rms $v_3$ as a function of $\bra p_t \ket$ in ultracentral U+U collisions (diamonds) and Au+Au collisions (circles) at top RHIC energy. The \trento{} implementation that yields these results is detailed in Appendix~\ref{app:tr}.}
    \label{fig:8}
\end{figure}

Figure~\ref{fig:8} shows my predictions for the $\bra p_t \ket$ dependence of triangular flow, $v_3$. As expected from the results of Pb+Pb collisions at LHC energy~\cite{Aad:2019fgl}, the correlation between $v_3$ and $\bra p_t \ket$ is positive in ultracentral collisions. Note that $v_3$ is slightly larger in U+U collisions, in agreement with STAR data~\cite{Adamczyk:2015obl}. Note that the comparison between this result and preliminary STAR data~\cite{shengli} is actually excellent. This result is very important. Since $v_3$ is not affected by the deformation parameter, obtaining a good description of data without any additional tuning provides a nontrivial confirmation of the goodness of the model implementation. I emphasize that the correlation between $v_3$ and $\bra p_t \ket$ predicted by my calculation would be perfectly flat, and so in stark disagreement with data, if I used $R$ as a predictor of $\bra p_t \ket$, instead of $E_0$, as shown explicitly in Ref.~\cite{Giacalone:2020dln}.

\begin{figure*}[t]
    \centering
    \includegraphics[width=.95\linewidth]{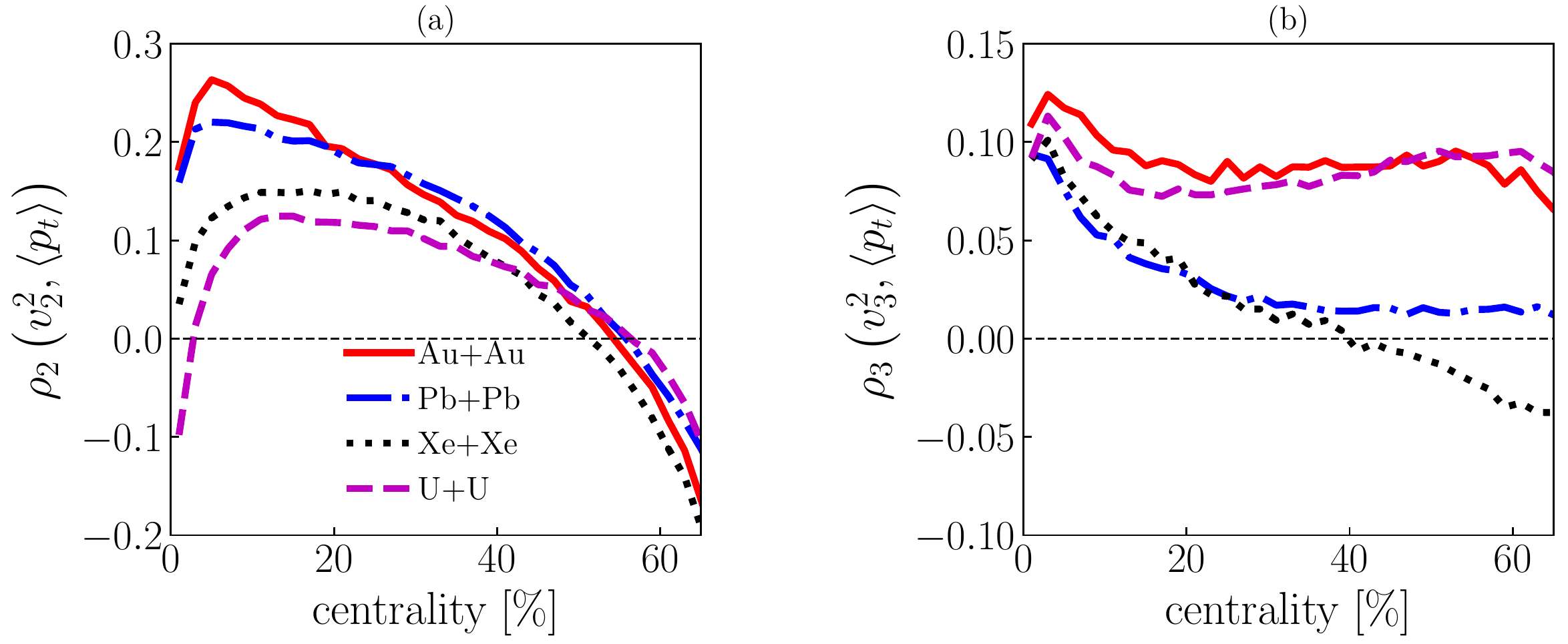}
    \caption{Predictions for $\rho_n\left ( v_n^2,\bra p_t \ket \right)$, as defined by Eq.~(\ref{eq:rhoe}), as a function of collision centrality for $^{197}$Au+$^{197}$Au and $^{238}$U+$^{238}$U collisions at RHIC energy, and $^{208}$Pb+$^{208}$Pb and $^{129}$Xe+$^{129}$Xe at LHC energy. (a) $\rho_2 \left ( v_2^2,\bra p_t \ket \right)$. (b) $\rho_3 \left ( v_3^2,\bra p_t \ket \right)$. These results are obtained with the \trento{} model (see Appendix~\ref{app:tr}). For Pb+Pb collisions, experimental data by the ATLAS collaboration are available in Ref.~\cite{Aad:2019fgl}, and were compared to the results of the same \trento{} calculation shown here in Ref.~\cite{Giacalone:2020dln}.}
    \label{fig:9}
\end{figure*}

\subsection{Removing statistical fluctuations: \\ Bo\.zek correlation coefficient}

\label{sec:42}

In analyses of heavy-ion collisions, one typically does not evaluate final-state observables, such as $\bra p_t \ket$ or $v_n$, on an event-by-event basis, because of the large statistical fluctuations associated with the finite number of particles, as explained in Sec.~\ref{sec:33}. Observables are typically obtained by statistical averages over many events, constructed in such a way to remove the trivial (and overwhelming) finite-$N$ effects.

For the correlation between $v_n$ and $\bra p_t \ket$, the corresponding observable was first studied in Ref.~\cite{Mazeliauskas:2015efa} in the context of a principal component analysis, and later by Bo\.zek~\cite{Bozek:2016yoj}, who formulated it as a simple correlation coefficient:
\begin{equation}
  \label{eq:rhon}
  \rho_n \left ( v_n^2, \bra p_t \ket \right ) \equiv \frac{\left\langle\langle p_t\rangle
    v_n^2\right\rangle
    -\left\langle\langle p_t\rangle\right\rangle\left\langle v_n^2\right\rangle}{\sigma_{p_t}\sigma_{v_n^2} }.
\end{equation}
where outer angular brackets denote an average over events in a given multiplicity (centrality) window. The standard deviations, $\sigma_{p_t}$ and $\sigma_{v_n}$, are given by:
\begin{align}
\label{sigmapt}
\nonumber \sigma_{p_t}&\equiv \sqrt{\left\langle \langle p_t\rangle^2\right\rangle
-\left\langle \langle p_t\rangle\right\rangle^2}, \\
\sigma_{v_n^2}&\equiv \sqrt{\left\langle v_n^4\right\rangle
-\left\langle v_n^2\right\rangle^2}.
\end{align}
In Ref.~\cite{Giacalone:2020dln}, it has been shown that if one replaces $v_n$ with $\varepsilon_n$, and $\bra p_t \ket$ with $E_0/S$, then the corresponding correlator:
\begin{equation}
  \label{eq:rhoe}
   \frac{\left\langle E_0/S
    \varepsilon_n^2\right\rangle
    -\left\langle E_0/S\right\rangle\left\langle \varepsilon_n^2\right\rangle}{
    \sigma_{E_0/S}    \sigma_{\varepsilon_n^2}
}  ,
\end{equation}
with an appropriate redefinition of the standard deviations in Eq.~(\ref{sigmapt}), provides a good description of experimental data on the correlation (\ref{eq:rhon}) measured in Pb+Pb collisions by the ATLAS collaboration~\cite{Aad:2019fgl}.

The Bo\.zek coefficient is extremely robust. Since it does not depend on the value of $\kappa_n$, it is essentially insensitive to the medium properties of the quark-gluon plasma, as confirmed to a great extent by both the results of Ref.~\cite{Giacalone:2020dln}, and the results shown in Fig.~9 of Ref.~\cite{Schenke:2020uqq}. Furthermore, as anticipated, the correlation coefficient in Eq.~(\ref{eq:rhon}) is by construction insensitive to trivial statistical fluctuations, and isolates the correlations coming from genuine collective effects. This implies that when I evaluate Eq.~(\ref{eq:rhoe}), I use directly the distribution of $E_0/S$ coming from the \trento{} model. I do not have to rescale the distribution by the factor $\kappa_0$ in Eq.~(\ref{eq:predictor}), nor to perform the correction described in Sec.~\ref{sec:33}.
\begin{figure*}[t]
    \centering
    \includegraphics[width=.9\linewidth]{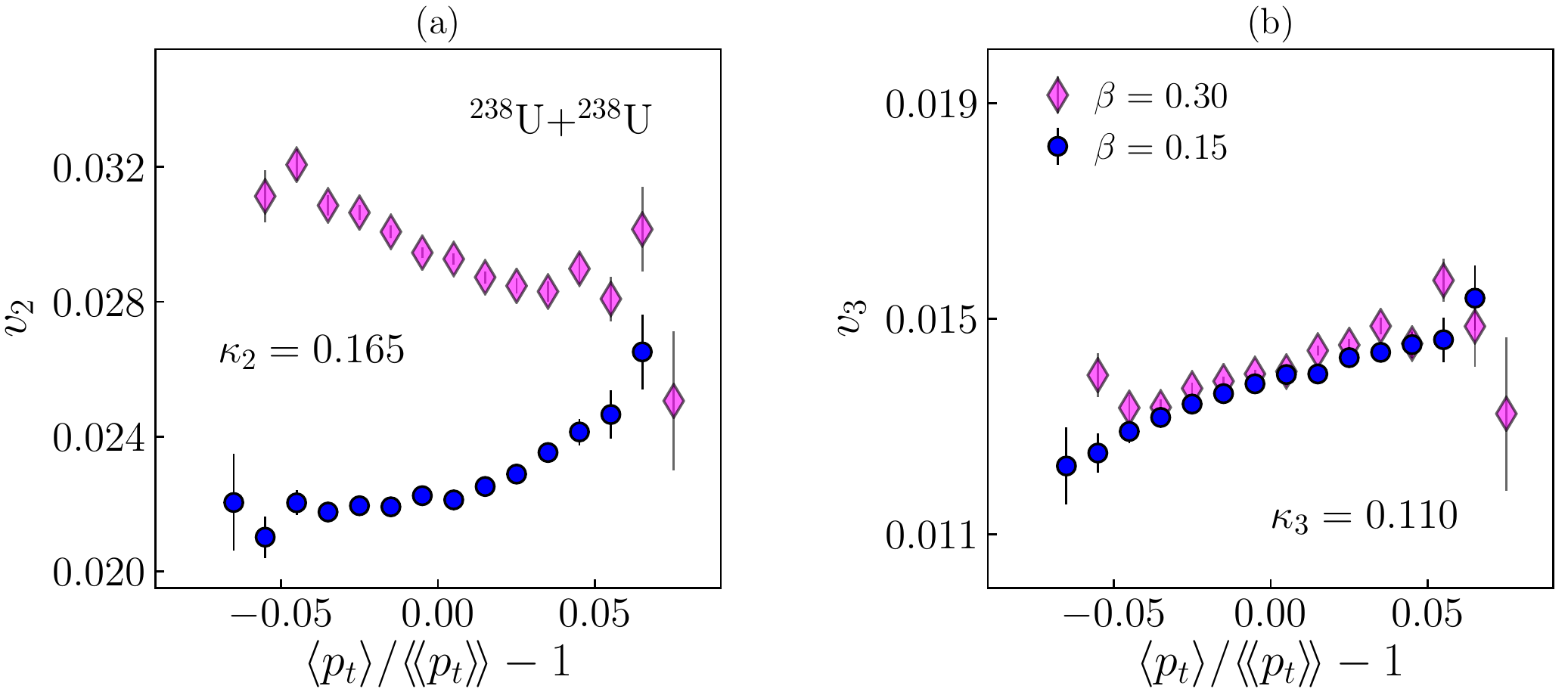}
    \caption{Rms anisotropic flow coefficients, $v_n$, as a function of $\bra p_t \ket$ in ultracentral U+U collisions at $\sqrt{s}=200$~GeV, for different values of the deformation parameter: $\beta=0.30$ [diamonds, corresponding to Fig.~\ref{fig:7}(a)], and $\beta=0.15$ (circles). (a) $v_2$. (b) $v_3$.}
    \label{fig:10}
\end{figure*}

I use Eq.~(\ref{eq:rhoe}) to make quantitative predictions for the Bo\.zek correlation coefficient in U+U and Au+Au collisions at top RHIC energy, as function of the collision centrality. The results are presented in Fig.~\ref{fig:9}. In panel (a), I observe that $\rho_2\left (v_2^2,\bra p_t\ket\right)$ is negative in central U+U collisions, while it is positive in central Au+Au collisions. This effect is compatible with Fig.~\ref{fig:7}, and is caused by the prolate deformation of $^{238}$U nuclei. Note that the value of $\rho_2\left (v_2^2,\bra p_t\ket\right)$ in U+U collisions is lower than in Au+Au collisions across essentially the full range of centrality. In panel (b) I show instead my predictions for $\rho_3\left (v_3^2,\bra p_t\ket\right)$. This quantity should not be sensitive to the value of $\beta_2$, and I observe, accordingly, only a minor difference between Au+Au and U+U.

It is insightful to look as well at the results for LHC systems, in particular, at the difference between collisions of spherical $^{208}$Pb nuclei and collisions of prolate $^{129}$Xe nuclei. In Fig.~\ref{fig:9}(a), we note that the comparison between Pb+Pb and Xe+Xe is similar to that found for the RHIC systems. The reason is clearly that $^{129}$Xe nuclei are prolate ($\beta=0.18$). Equally striking, though, is the result shown in Fig.~\ref{fig:9}(b), where we observe now four curves that fall essentially into two distinct categories. RHIC systems present $\rho_3\left (v_3^2,\bra p_t\ket\right)$ of order 0.1 and mildly dependent on centrality, while LHC systems present a steep decrease with the centrality percentile. This is a remarkable byproduct of this analysis. The feature of the \trento{} model that drives this result is the fluctuation parameter, $k$, discussed in Appendix~\ref{app:tr}. This quantity determines the magnitude of initial-state fluctuation, and is here tuned in such a way that fluctuations at RHIC energy are essentially twice as large as fluctuations at LHC energy (as explained in Appendix~\ref{app:tr}). Therefore, my results suggest that $\rho_3\left (v_3^2,\bra p_t\ket\right)$ could serve as a fine probe of the evolution of fluctuations from RHIC energy to LHC energy. 
\begin{figure*}[t]
    \centering
    \includegraphics[width=.93\linewidth]{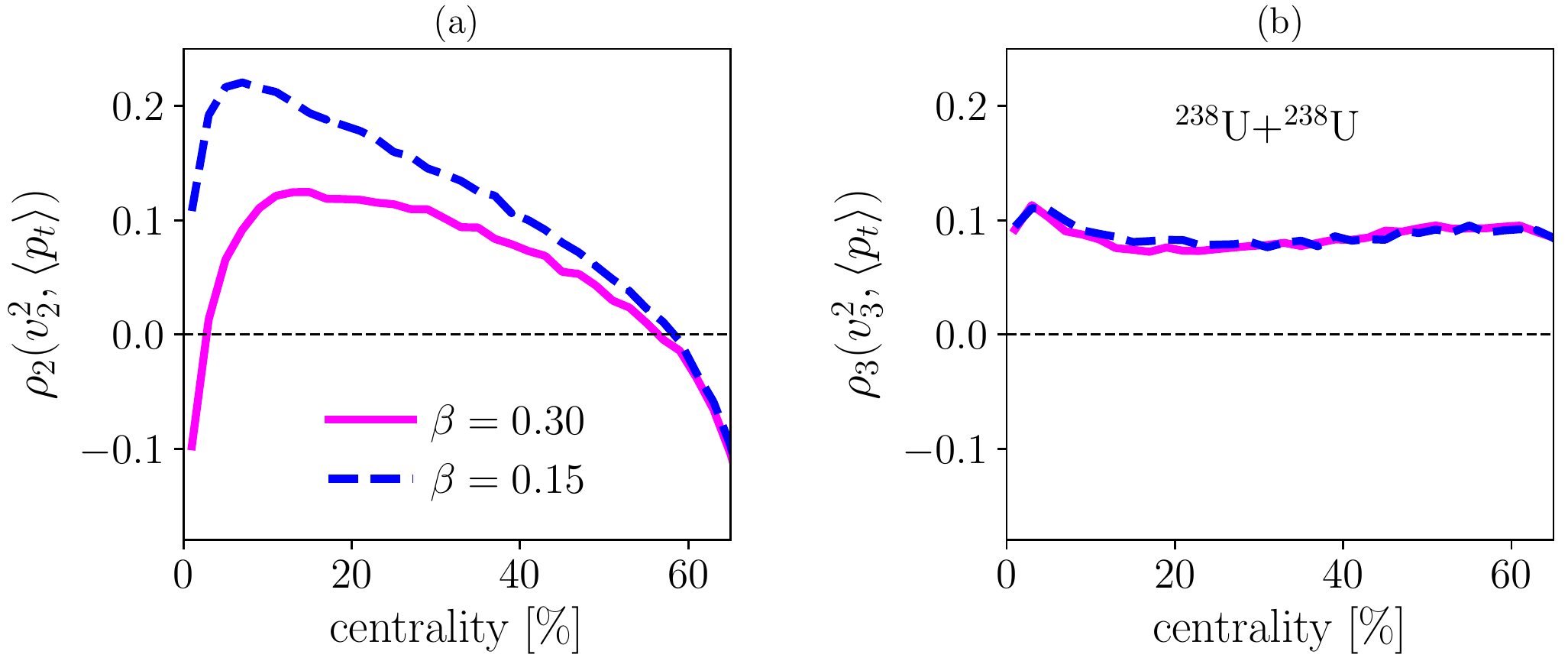}
    \caption{Predictions for the Bo\.zek correlation coefficient, $\rho_n\left ( v_n^2,\bra p_t \ket \right)$, as a function of collision centrality for U+U collisions at top RHIC energy, utilizing two distinct values of $\beta$. (a) $n=2$. (b) $n=3$. Solid lines: $\beta=0.3$. Dashed lines: $\beta=0.15$.}
    \label{fig:11}
\end{figure*}

\subsection{Role of the quadrupole deformation}

\label{sec:43}

My claim in this paper is that future STAR data will allow us to place a constraint on the value of $\beta$. To check this, I assess the role of $\beta$ by repeating the previous calculations with a quadrupole deformation reduced by a factor 2, i.e. $\beta=0.15$, in the colliding $^{238}$U nuclei. All the other model parameters are kept fixed.\footnote{Note that, while varying $\beta$, one should  make sure that the resulting parametrization of the nuclear density remains consistent with the observed transition probability of the electric quadrupole operator, as explained in Ref.~\cite{Shou:2014eya}, and otherwise retune the other shape parameters, $a$ and $R$, to meet this condition. For the parametrizations used in this paper, I have not performed any such checks. However, this issue will have to be taken into proper account in future quantitative extractions of the value of $\beta$ from experimental data.}

The resulting flow coefficients as function of $\bra p_t \ket$ are shown as full circles in Fig.~\ref{fig:10}. Panel~(a) may be considered as the most important result of this paper. I find that, with $\beta=0.15$, the correlation between $\bra p_t \ket$ and $v_2$ is positive, and the effect of the deformation is essentially not visible. Panel~(b) shows, on the other hand,  that modifying $\beta$ does not affect $v_3$.  In Fig.~\ref{fig:11}, I show as well the correlation coefficient of Bo\.zek for two values of $\beta$. The result for $\rho_2\left ( v_2^2, \bra p_t \ket \right)$ is in panel~(a). The sensitivity of this observable to the value of the quadrupole parameter is impressive. Differences between the curves corresponding to $\beta=0.15$ and $\beta=0.3$ are visible up to 40\% centrality, whereas usually the effects of deformation on quantities such as the cumulants of elliptic flow are typically inferrable only from central collisions~\cite{Acharya:2018ihu}. In panel~(b) I show results for $\rho_3\left ( v_3^2, \bra p_t \ket \right)$. As expected, the two curves overlap.

Let me stress once more that this calculation has no free parameters. The most uncertain ingredient in Fig.~\ref{fig:10}(a) is the value of $\kappa_2$, but this quantity changes only the normalization of the results, and not their shape. The results in Fig.~\ref{fig:10} suggest then that it would not be possible for state-of-the-art simulations of heavy-ion collision to get a reasonable description of preliminary STAR data~\cite{shengli} without implementing a value of $\beta$ close to 0.3. My conclusion is that heavy-ion collision data are in agreement with the result of Raman \textit{et. al}~\cite{Raman:1201zz}, and that $^{238}$U nuclei have indeed a quadrupole deformation parameter of order 0.3. Future investigations of $\rho_2\left ( v_2^2, \bra p_t \ket \right )$ at RHIC will further support (or refute) this conclusion.

\section{Conclusion and outlook}

\label{sec:5}

The observation of an anticorrelation between $\bra p_t \ket$ and $v_2$ in ultracentral U+U collisions provides, arguably, the most striking signature of nuclear deformation so far observed in high-energy nuclear experiments, and should be regarded as compelling evidence of a significant prolate deformation, $\beta\approx0.3$, in $^{238}$U nuclei.

The ingredient missing in the qualitative analysis of Ref.~\cite{Giacalone:2019pca} is a realistic predictor for the event-by-event $\bra p_t \ket$. The present works fixes this problem by implementing a predictor based on the initial energy, which in turn allows me to carry out quantitative predictions for the phenomenon observed in preliminary STAR data. For large nuclei with a significant prolate deformation, such as $^{238}$U, spectacular results can be obtained by measuring $v_2$ as a function of $\bra p_t \ket$ [Fig.~\ref{fig:7}(a)], although one has to carefully consider the impact of statistical fluctuations.  The Bo\.zek correlation coefficient [Fig.~\ref{fig:9}(a)] is instead insensitive to trivial finite-$N$ effects, and also largely independent of the details of the hydrodynamic modeling. At the same time, it displays an outstanding sensitivity to the value of $\beta$ [Fig.~\ref{fig:11}(a)], and, therefore, it should be investigated at RHIC, and in Xe+Xe collisions at LHC.

My conclusion is that, if we look at observables that have a great sensitivity to the quadrupole deformation of the colliding nuclei, then we can use relativistic nuclear collisions to constrain the value of $\beta$. The correlation between $v_2$ and $\bra p_t \ket$ provides such an observable. 

The realization of Ref.~\cite{Giacalone:2019pca} concerning the role of $\bra p_t \ket$ for the phenomenology of nuclear deformation opens a new direction for studies of nuclear structure at high-energy nuclear colliders. I expect more observables based on $\bra p_t \ket$, and sensitive to the deformation of the colliding nuclei, to be invented in the near future. One should also consider the possibility of performing new experiments with new species. Heavy-ion collisions could be used to provide independent confirmations of the results of low-energy experiments, as well as to put constraints on the value of $\beta$ for nuclei that have not yet been investigated experimentally. Such studies would be feasible thanks to the great versatility of the RHIC machine.

\section{Acknowledgments}

I thank Shengli Huang and Prithwish Tribedy for their continued interest, and Jean-Yves Ollitrault for the careful reading of the manuscript. Useful discussions with Thomas Duguet, Fernando Gardim, Jiangyong Jia, Matt Luzum, Aleksas Mazeliauskas, Jean-Yves Ollitrault, Bj\"orn Schenke, Vittorio Som\`a, and Derek Teaney are kindly acknowledged.

\appendix

\section{The \trento{} model}

\label{app:tr}

The \trento{} model~\cite{Moreland:2014oya} is a flexible generalization of the Glauber Monte Carlo model~\cite{Miller:2007ri}. The modeling of the collision process starts with a random sampling of nucleons within the nuclear volume. The coordinates of the nucleons are generated according to the 2-parameter Fermi density shown in Eq.~(\ref{eq:density}):
\begin{equation}
\label{eq:ws}
\rho(x,y,z) = \frac{\rho_0}{1 + \exp \biggl[ \frac{1}{a} \biggl( r-R_0 \bigl( 1+\beta_2 Y_{20} \bigr) \biggr) \biggr]},
\end{equation}
where $r=\sqrt{x^2+y^2+z^2}$, $a$ and $R_0$ are the diffusiveness and radius of the nucleus, respectively, the spherical harmonic $Y_{20}=\sqrt[]{\frac{5}{16 \pi}}(3\cos^2 \Theta-1)$ induces a dependence of the density on the solid angle $\Theta$ in the intrinsic nuclear frame, while the parameter $\beta$ quantifies the magnitude of the ellipsoidal deformation. The parameters used to model the nuclei considered in this paper are reported in Tab.~\ref{tab:1} at the end of this appendix.

Once the transverse coordinates, $(x,y)$, of the nucleons are known, the nuclei are overlapped at a random impact parameter, and nucleon-nucleon collisions take place. A participant nucleon carries a Gaussian profile of matter density, $\rho_s({\bf x})$, in the transverse plane:
\begin{equation}
\label{eq:gauss}
\rho_{s,i} ({\bf x})= \frac{\omega_i}{2\pi \sigma^2} \exp \biggl [ -\frac{({\bf x}-{\bf x}_i)^2}{2\sigma^2} \biggr],
\end{equation}
where the index $i$ labels the $i$-th participant nucleon, and I use $\sigma=0.5$~fm.
The weight of each participant, $\omega_i$, is random and follows a gamma distribution of unit mean:
\begin{equation}
P(\omega) = \frac{k^k \omega^{k-1}e^{-k}}{\Gamma(k)}.
\end{equation}
Note that the variance of this distribution is equal to $k^{-1}$.

The total density in a given nucleus, say A, is given by
\begin{equation}
s_A ({\bf x}) = \sum_i \rho_{s,i}({\bf x}),
\end{equation}
while the total entropy density profile for two colliding nuclei is proportional to a generalized mean:
\begin{equation}
s({\bf x}) \propto \biggl ( \frac{s_A({\bf x})^p + s_B({\bf x})^p}{2} \biggr )^{1/p},
\end{equation}
where $p$ is any real number. I use $p=0$, which is the value favored by the comprehensive Bayesian analyses of Refs.~\cite{Moreland:2018gsh,Bernhard:2019bmu}, corresponding to a geometric mean:
\begin{align}
\nonumber s({\bf x}, \tau_0) &= \frac{N_0}{\tau_0} \biggl ( \frac{s_A({\bf x})^p + s_B({\bf x})^p}{2} \biggr )^{1/p}\biggl|_{p=0} \\ 
&= \frac{N_0}{\tau_0} \sqrt[]{s_A({\bf x})s_B({\bf x})},
\end{align}
where $\tau_0$ is the initial time, while the constant $N_0$ is added in order for the model to yield the right final-state multiplicity at the end of hydrodynamics. Hydrodynamic simulations are presented in Appendix~\ref{app:A}, where I implement $\tau_0=0.2$~fm/$c$, $N_0=21.6$ to reproduce the charged-particle multiplicity measured at top RHIC energy. Note that the $p=0$ Ansatz for the entropy density yields a correlation between anisotropy and multiplicity at fixed number of participant nucleons that is consistent with that reported by the STAR collaboration in ultracentral U+U collisions~\cite{Moreland:2014oya,Adamczyk:2015obl}, thus making it a naturally viable prescription for modeling collisions of deformed nuclei.

I discuss now the choice of the fluctuation parameter, $k$. Substantial experimental evidence, coming from analyses of anisotropic flow fluctuations, multiplicity fluctuations, and dynamical $\bra p_t \ket$ fluctuations, points to the fact that initial-state fluctuations are larger at RHIC energy than at LHC energy (see e.g. Ref.~\cite{Giacalone:2019vwh} for a short review). I follow, accordingly, the phenomenological applications of Refs.\cite{Giacalone:2017dud,Giacalone:2018apa}, and implement $k=0.5$ for collisions at RHIC energy, and $k=2$ for collisions at LHC energy. This implies that, in my model, initial-state fluctuations at RHIC energy are essentially larger than at LHC energy by a factor 2.

Finally, the \trento{} model is used to sort events into centrality classes. I generate about $10^7$ events for each collision system, and evaluate the corresponding distribution of the total entropy per unit rapidity, $S=\tau_0 \int_{\bf x} s({\bf x},\tau_0)$. Events are then classified according to their value of $S$ to mimic the centrality selection performed in experiment, where events are classified according to their final-state multiplicity~\cite{Abelev:2013qoq,Aaboud:2019sma}. Ultracentral events at top RHIC energy have typically $S\sim4000$ at midrapidity. The class defining ultracentral events in this paper, 0.4-0.8\%, corresponds to $4072 < S < 4231$ for U+U collisions, and to $3419<S<3559$ for Au+Au collisions. The hydrodynamic simulations of U+U collisions performed in Appendix~\ref{app:A} evolve events taken from the 0.78-0.96\% class, which corresponds to $4023 < S < 4078$. 
\begin{table}[H]
\centering
\begin{tabular}{|c|c|c|c|}
\hline
species& $a$ [fm] & $R$ [fm] & $\beta$ \cr
\hline
$^{238}$U~\cite{DeJager:1987qc} & 0.60 & 6.80 & 0.30\cr
$^{208}$Pb~\cite{DeJager:1987qc} & 0.55 & 6.62 & 0 \cr
$^{197}$Au~\cite{DeJager:1987qc} & 0.53 & 6.40 & -0.13~\cite{Moller:2015fba,Hilaire:2007tmk} \cr
$^{129}$Xe~\cite{Acharya:2018hhy} & 0.59 & 5.40 & 0.18~\cite{Acharya:2018hhy}  \cr
\hline
\end{tabular}
\caption{\label{tab:1} 
Parameters used in Eq.~(\ref{eq:density}) for different species.
}
\end{table}
\begin{figure*}[t]
    \centering
    \includegraphics[width=.95\linewidth]{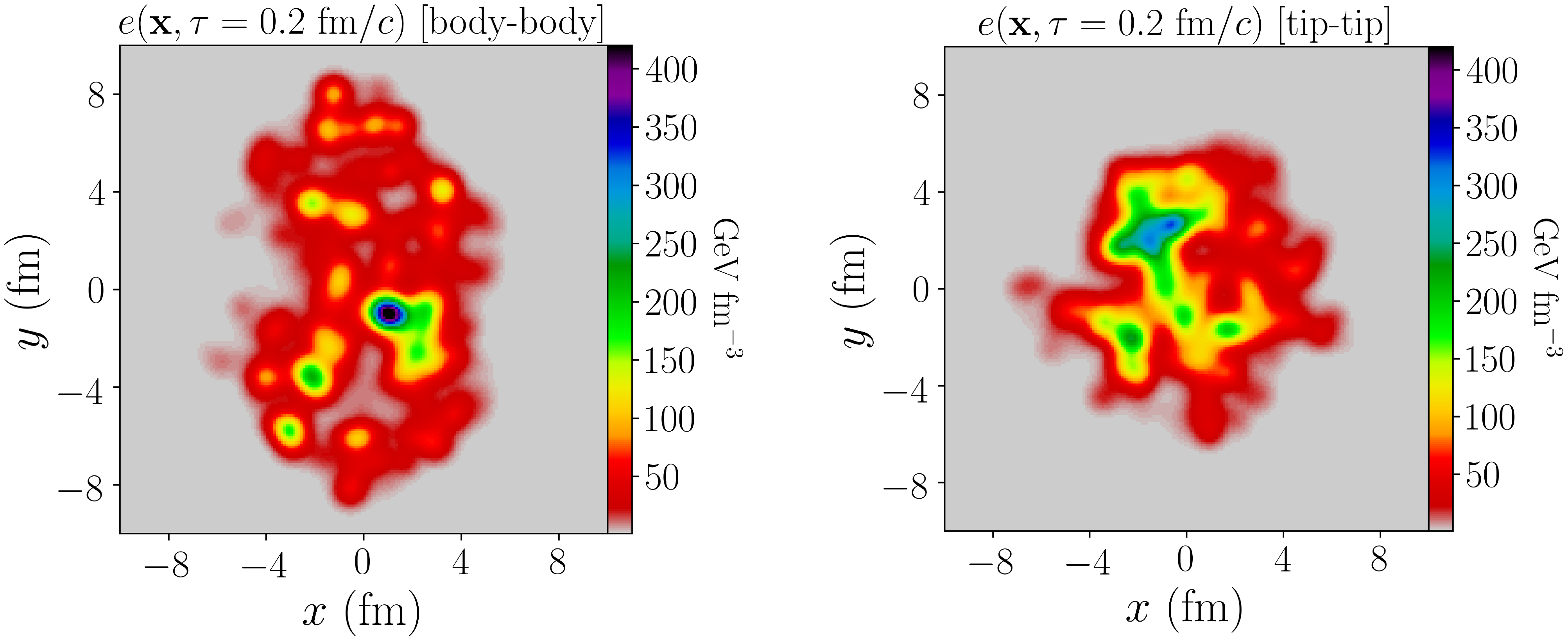}
    \caption{Left: initial energy density profile for a body-body collision, with $S=4040$, $E_0=1294$~GeV, $\varepsilon_2=0.478$, $\varepsilon_3=0.191$, and average temperature $\bra T \ket = 0.433$~GeV. At the end of hydrodynamics, this event yields: $dN_{ch}/d\eta_{|\eta|<1}=1296$, $\bra p_t \ket=0.587$~GeV, $v_2=0.083$, $v_3=0.016$, where the flow coefficients are obtained with the kinematic cuts of the STAR collaboration: $|\eta|<1$, and $0.2<p_t<2$~GeV. Right: a tip-tip collision, with $S=4072$, $E_0=1429$~GeV, $\varepsilon_2=0.096$, $\varepsilon_3=0.089$, and average temperature $\bra T \ket = 0.475$~GeV. This event yields $dN_{ch}/d\eta_{|\eta|<1}=1280$, $\bra p_t \ket=0.651$~GeV, $v_2=0.027$, $v_3=0.009$.}
    \label{fig:12}
\end{figure*}

\section{Hydrodynamic simulations}

\label{app:A}

In this section I perform hydrodynamic simulations of central U+U collisions with a threefold purpose:
\begin{itemize}
    \item Confirm that $\bra p_t \ket$ is strongly correlated with $E_0$ in viscous hydrodynamics.
    \item Confirm that body-body collisions yield lower $\bra p_t \ket$ than tip-tip collisions.
    \item Confirm that the response coefficients $\kappa_n$ used in the phenomenological applications of this paper [see Eq.~(\ref{eq:kappan2})] are in agreement with the results of full hydrodynamic simulations.
\end{itemize}

To do this, I evolve hydrodynamically 60 profiles of entropy density coming from the \trento{} simulations described in Appendix~\ref{app:tr}. The calculation is carried out for both both body-body and tip-tip configurations. The initial profiles are selected in a narrow bin of entropy corresponding to the 0.78--0.96\% centrality class, as explained in Appendix~\ref{app:tr}. 

It is instructive to have a look at an example of initial density profile for both a body-body and a tip-tip event. In the left panel of Fig.~\ref{fig:12}, I show the initial energy density profile, $e({\bf x},\tau_0)$, for a body-body collision. On the right panel, I show instead the energy density profile of a tip-tip collision. One immediately observes the different global geometry of these profiles, with an enhanced elliptical asymmetry in the body-body event. Secondly, we can infer essentially by eye that the tip-tip collision covers a smaller area in the transverse plane, and presents on average a larger value of energy density (or temperature). This is the reason why the tip-tip event yields a larger value of $\bra p_t \ket$ at the end of the hydrodynamic expansion. 
\begin{figure}[b]
    \centering
    \includegraphics[width=\linewidth]{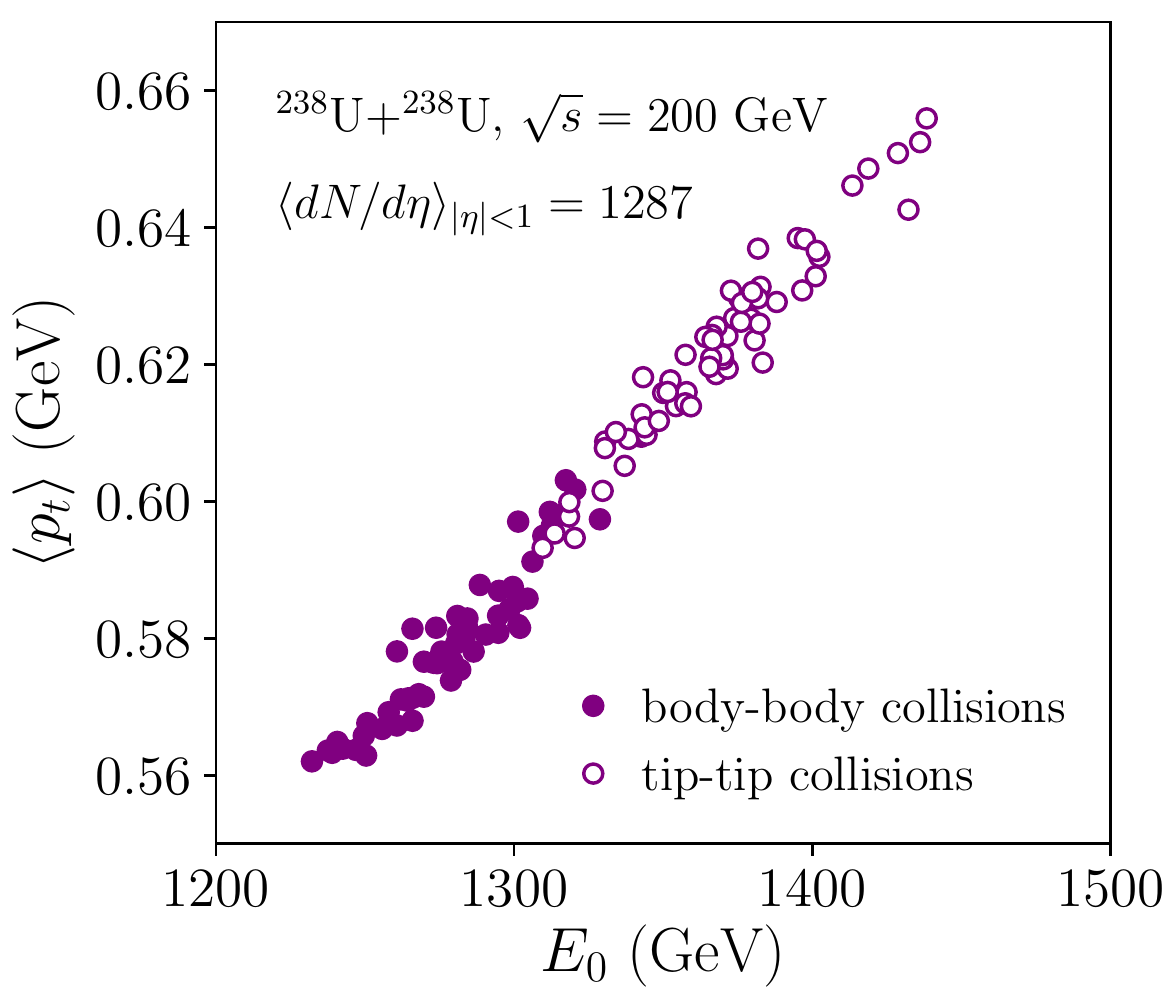}
    \caption{Average transverse momentum, $\bra p_t \ket$ as a function of the initial energy, $E_0$, in hydrodynamic simulations of central U+U collisions at top RHIC energy. The charged-particle multiplicity in the $|\eta| < 1$ window corresponds approximately to 1\% centrality in the STAR analysis. Full symbols: body-body collisions. Empty symbols: tip-tip collisions.}
    \label{fig:13}
\end{figure}
\begin{figure*}[t]
    \centering
    \includegraphics[width=.95\linewidth]{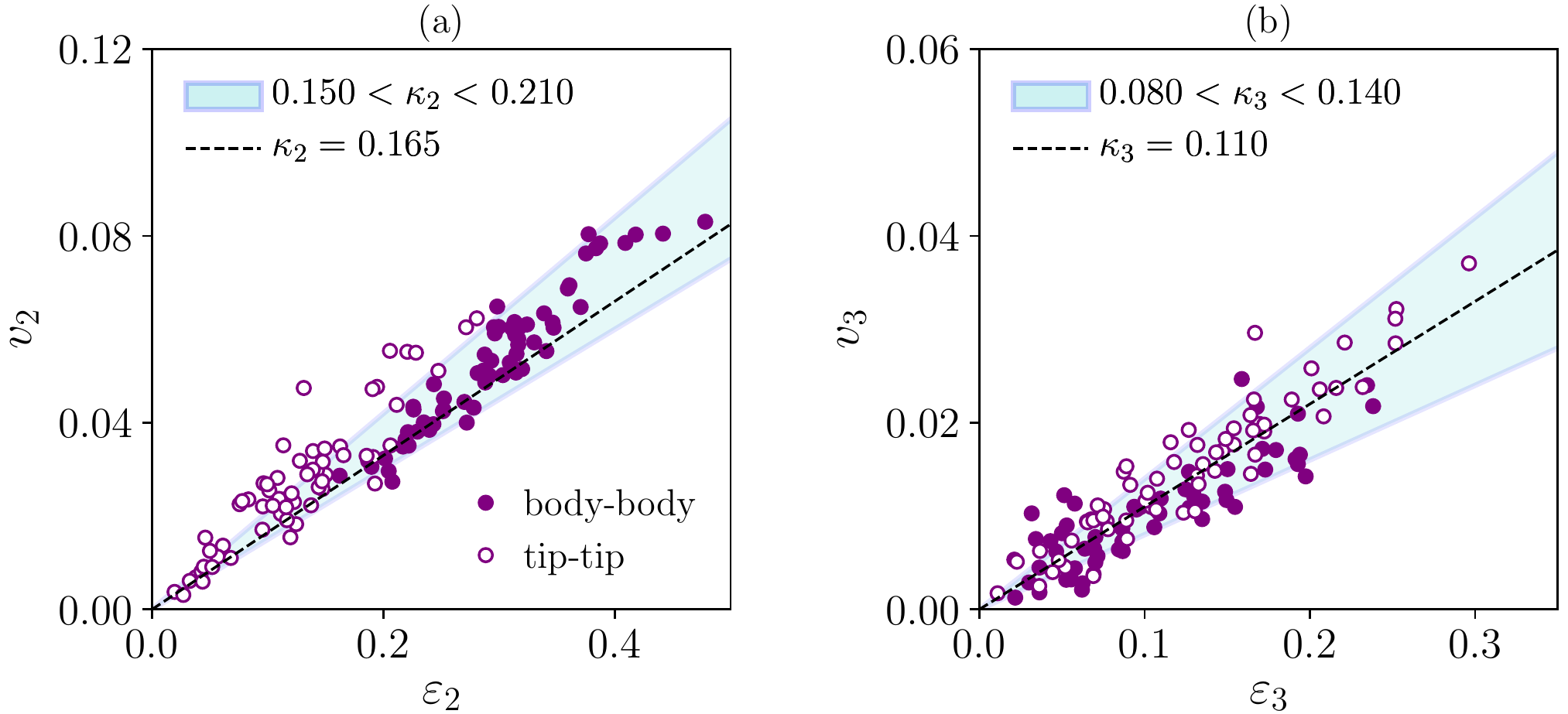}
    \caption{Flow coefficients, $v_n$, as a function of the initial spatial anisotropy, $\varepsilon_n$, in high-multiplicity U+U collisions. The flow coefficients are evaluated within kinematic cuts $|\eta|<1$ and $0.2 < p_t < 2$~GeV, following the STAR collaboration. Shaded bands are meant to provide the range of viable values for the response coefficients $\kappa_n=v_n/\varepsilon_n$. Dashed lines correspond to the values implemented in Fig.~\ref{fig:7} and in Fig.~\ref{fig:8}. Left panel: elliptic flow, $v_2$. Right panel: triangular flow, $v_3$.}
    \label{fig:14}
\end{figure*}

The boost-invariant hydrodynamic evolution of the initial profiles is carried out by means of the MUSIC hydrodynamic code~\cite{Schenke:2010nt,Schenke:2010rr,Paquet:2015lta}. In my setup, I neglect the pre-equilibrium phase~\cite{Kurkela:2018wud,Schlichting:2019abc}, and I start hydrodynamics at initial time $\tau_0 = 0.2$~fm/$c$. The hydrodynamic expansion is viscous. I implement a constant value of the specific shear viscosity, $\eta/s=0.16$, and a temperature-dependent specific bulk viscosity, $\zeta/s$. The current status of the implementation of $\zeta/s$ in heavy-ion collisions is fairly uncertain~\cite{Byres:2019xld}. On the one hand, calculations starting with IP-Glasma~\cite{Schenke:2012wb} initial conditions implement a bulk viscosity that peaks around $\zeta/s\approx0.3$, to yield values of $\bra p_t \ket$ in agreement with experimental data, at both RHIC and LHC energy~\cite{McDonald:2016vlt,Ryu:2017qzn,Schenke:2019ruo}. On the other hand, calculations starting with \trento{} initial condition are able to fit data on $\bra p_t \ket$ with a $\zeta/s$ that is smaller by essentially one order of magnitude ($\zeta/s\approx0.03$ at the peak~\cite{Moreland:2018gsh,Bernhard:2019bmu}), although results are available only at LHC energy. Here I work at RHIC energy starting with \trento{} initial conditions, so I set up a sort of hybrid scenario: I implement a bulk viscosity having the same functional form as in Refs.~\cite{Ryu:2015vwa,Ryu:2017qzn}, but I reduce the value of $\zeta/s$ at its peak by a factor 10. The shear and bulk corrections to the momentum distribution functions at freezeout are the same as in Ref.~\cite{Ryu:2017qzn}.

 The medium has the equation of state of lattice QCD~\cite{Borsanyi:2013bia}, and fluid cells convert into hadrons when their temperature falls below $T=0.15$~GeV. At freezeout, all hadronic resonances can be formed~\cite{Broniowski:2001we,Alba:2017hhe}, and their decays~\cite{Mazeliauskas:2018irt} to stable hadrons are taken into account. The outcome is a final boost-invariant spectrum of charged hadrons, $\frac{dN}{d^2{\bf p}_t}$, which is then integrated to calculate the charged-particle multiplicity, the average transverse momentum, and the flow coefficients, see e.g. Eq.~(\ref{eq:1}) and Eq.~(\ref{eq:flow}).

I start by showing the values of $\bra p_t \ket$ as a function of $E_0$ in Fig.~\ref{fig:13}. We observe the anticipated nearly-one-to-one correspondence between these quantities. The correlation is essentially as strong as that observed in Refs.~\cite{Gardim:2020sma,Giacalone:2020dln}, meaning that the simple picture of $\bra p_t \ket$ as a measure of the energy per particle in the initial state is only mildly disrupted by the viscous corrections, which are in fact large in my hydrodynamic setup. The second remarkable feature displayed by the results of Fig.~\ref{fig:13} is that body-body and tip-tip events, though falling essentially on the same curve, cover distinct intervals in $E_0$ and $\bra p_t \ket$, which confirms the overall picture of this paper, i.e., that $\bra p_t \ket$ provides a powerful handle on the orientation of the colliding ellipsoids. Note that the correlation in Fig.~\ref{fig:13} would be even stronger if one replaced $E_0$ with $E_0/S$, which is in fact the predictor used in this paper.

Finally, I calculate flow coefficients, $v_n$, that I plot as functions of $\varepsilon_n$ in Fig.~\ref{fig:14}. 

Figure~\ref{fig:14}(a) shows results for elliptic flow. The first notable feature is that, as expected, body-body and tip-tip collision cover essentially two distinct intervals in $\varepsilon_2$ and $v_2$, as discussed in Sec.~\ref{sec:32}. The shaded band shows a range of viable values for the response coefficient $\kappa_2 = v_2 / \varepsilon_2$. The dashed line corresponds to $\kappa_2=0.165$, which is the value suggested in Ref.~\cite{Giacalone:2018apa}, and used here in Figs.~\ref{fig:7}~and~\ref{fig:10}. One sees that $\kappa_2=0.165$ captures a significant fraction of the body-body points, but only a minor fraction of tip-tip events, meaning that the prediction shown in Fig.~\ref{fig:7} is smaller than the full hydrodynamic result by about 10\%. Note also that, contrary to Fig.~\ref{fig:13}, body-body and tip-tip collisions do not fall on the same curve. The response coefficient is slightly larger, by a few percents, in the tip-tip case. This is consistent with previous studies showing that the linear response coefficient decreases when the initial-state anisotropy is large~\cite{Noronha-Hostler:2015dbi,Sievert:2019zjr}, as it occurs, e.g., moving from central to peripheral nucleus-nucleus collisions. The coefficient $\kappa_2$ presents, as a consequence, a small dependence on $\bra p_t \ket$, which should however be negligible in Fig.~\ref{fig:7} thanks to the strong smearing due to statistical fluctuations.
\begin{figure*}[t]
    \centering
    \includegraphics[width=.92\linewidth]{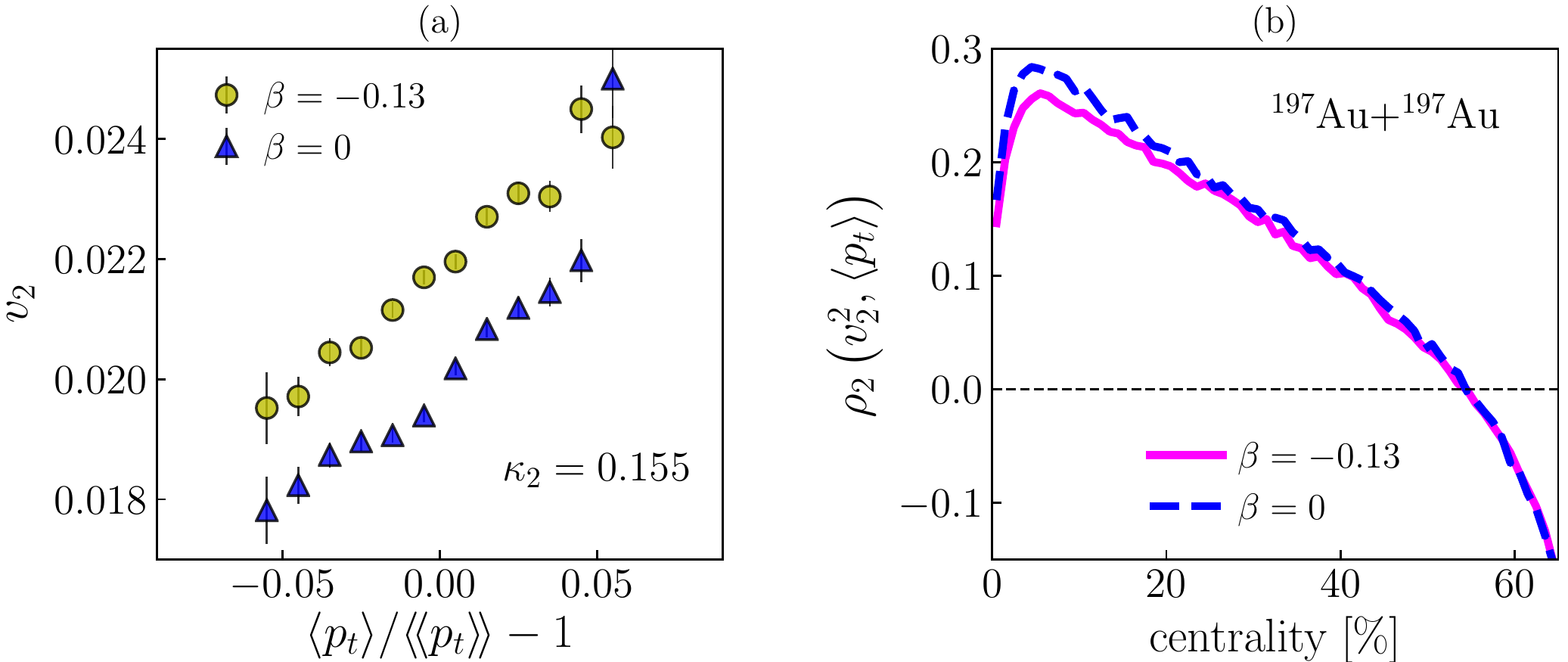}
    \caption{Left: $v_2$ as a function of the relative variation of $\bra p_t \ket$ in ultracentral Au+Au collisions at top RHIC energy. Full circles are obtained with oblate $^{197}$Au nuclei ($\beta=-0.13$), while triangles are obtained with spherical nuclei. Right: Bo\.zek coefficient, $\rho_2\left ( v_n^2, \bra p_t \ket \right)$. Solid line: $\beta=-0.13$. Dashed line: $\beta=0$. }
    \label{fig:15}
\end{figure*}

Figure~\ref{fig:14}(b) shows, on the other hand, results for triangular flow. Body-body and tip-tip collisions overlap to a good extent, although both $\varepsilon_3$ and $v_3$ are larger in tip-tip events. The value of the response coefficient chosen in Figs.~\ref{fig:8}~and~\ref{fig:10}, $\kappa_3=0.110$, provides a good description of the relation between $\varepsilon_3$ and $v_3$ observed in Fig.~\ref{fig:14}.

A final word about the response coefficients, $\kappa_2$ and $\kappa_3$, employed in Fig.~\ref{fig:7} and in Fig.~\ref{fig:8} for Au+Au collisions. I do not estimate them by means of hydrodynamic simulations. The value $\kappa_2=0.155$ is suggested by Ref.~\cite{Giacalone:2018apa}, where it is used to match the second-order cumulant of the eccentricity, $\varepsilon_2\{2\}$, computed in the same \trento{} model used here, to STAR data on the second-order cumulant of elliptic flow, $v_2\{2\}$, in ultracentral Au+Au collisions. I play the same game for the third harmonic. In STAR ultracentral data~\cite{Adamczyk:2015obl}, $v_3\{2\}=0.0138$, while my \trento{} calculation yields $\varepsilon_3\{2\}=0.1330$, hence $\kappa_3\simeq0.100$. Note that these coefficients are both smaller than those implemented in U+U collisions. This is consistent with the fact that $\kappa_n$ is damped by viscous corrections, which are larger in Au+Au collisions due to the smaller system size.

\section{Effect of oblate deformation in $^{197}$Au}

\label{app:B}

I assess the effect of $\beta$ on the the correlation between $v_2$ and $\bra p_t \ket$ in Au+Au collisions. This analysis is motivated by Ref.~\cite{Giacalone:2018apa}, where it was found that a deformation parameter $\beta=-0.13$ in the colliding $^{197}$Au nuclei is at variance with the experimental result that the fourth-order cumulant of elliptic flow, $v_2\{4\}^4$, is negative in central Au+Au collisions~\cite{Adamczyk:2015obl}. It is relevant, hence, to figure out whether $\beta$ plays any role for the observable studied in this paper.

I look at the correlation between $\bra p_t \ket$ and $v_2$, implementing both oblate and spherical $^{197}$Au nuclei. Figure~\ref{fig:15}(a) shows $v_2=0.155\varepsilon_2$ as a function of $\bra p_t \ket$. As expected, $v_2$ is smaller in magnitude in the calculation implementing spherical nuclei. However, the slope of the curve is just as large as with deformed nuclei. The effect of $\beta$ can be simply accounted, then, by a rescaling of the value of $\kappa_2$ by few percents. This is also confirmed in Fig.~\ref{fig:15}(b), where the Bo\.zek coefficient, independent of $\kappa_2$, turns out to be insensitive to the value of $\beta$.

\end{document}